\documentclass[aps,prstab,superscriptaddress,longbibliography, amsmath,amssymb]{revtex4-2}
\usepackage{amssymb,amsmath}
\usepackage{graphicx}
\usepackage{hyperref}
\hypersetup{
    colorlinks=true,
    linkcolor=blue,
    citecolor=blue,
    filecolor=magenta,
    urlcolor=blue
}

\begin{document}

\title{Wakefield Acceleration in a Layered Plasma Waveguide}


\author{G. V. Sotnikov}
\author{K. V. Galaydych}
\author{P. I. Markov}

\address{National Science Center Kharkiv Institute of Physics and Technology \\
1, Akademichna St., Kharkiv, 61108, Ukraine}



\begin{abstract}
Plasma wakefield accelerators (PWFA) represent one of the promising new accelerator concepts that are now being developed intensively for future applications in high-energy physics and industry.  Among the unresolved problems of practical implementation of PWFA there are maintaining the required quality (emittance, size, energy spread of bunches) and stable transportation of drive and accelerated (witness) bunches over long acceleration distances. For improving the bunch transport, we propose in this work that the plasma channel method should be modified through filling the bunch transport channel with the background plasma, the density of which is lower than the main plasma density building up an accelerating wake wave. We call this waveguide structure with a non-uniform transverse plasma profile a layered plasma waveguide (LPW). The wakefield excitation by a regular sequence of electron bunches in the LPW of cylindrical configuration has been explored both  analytically and numerically. The layered plasma has been modeled as a combination of a tubular plasma and a plasma column of significantly different densities. The plasma column has a lower density. In the linear approximation of plasma dynamics, analytical expressions for the excited longitudinal and transverse wakefields were obtained. The dispersion dependencies of the TM-modes of the LPW was obtained and analyzed, and it was found that there  was a single TM wave resonant with the electron bunch. Based on the obtained analytical expressions, the structures of the axial and radial wakefield amplitudes have been numerically investigated for the cases of a single drive bunch and  a regular train of bunches. It is shown that for certain density  ratios   of the outer and inner plasmas, it is possible   both to accelerate and focus simultaneously the drive and witness bunches. A spectral amplitude-frequency analysis of the excited wakefields has been carried out.
The 2.5D particle-in-cell code was used to simulate the witness acceleration of an electron bunch by a wakefield created by drive electron bunches in a two-layer plasma wakefield accelerator. The simulation showed good agreement with the results of analytical calculations.
\end{abstract}

\maketitle

\section{Introduction}\label{sec:1}
The beam-driven plasma wakefield accelerators (PWFA), which are one of the promising trend of new accelerator concepts~\cite{Chen_PRL1985, Rosenzweig_PRL1988}, are currently being intensively developed for future applying in high-energy physics and industry~\cite{Bingham_PPCF2004, Albert_2021,Gessner:2025acq}. They are able to provide very high and extremely high rates of acceleration of charged bunches, which makes them attractive for the use in future electron/positron colliders.  Despite the already demonstrated capability of producing high-intensity wakefields and the energy gain by witness bunches in these fields~\cite{Litos_2014,Litos_2016,Vafaei-Najafabadi_2016,CHIADRONI2017139,Joshi_2018, Lindstrrøm_PRL2021}, the challenges that are the subject of intensive research, viz., preserving emittance and required size of bunches as they move along the accelerating structure~\cite{Lindstrøm_PRL2024}, have not been fully solved.  In the simplest case with axial injection of bunches, the task goes over into transportation of bunches  through the channel of predetermined transverse dimensions. In the case of off-axis beams, PWFAs face another additional issue, namely, their susceptibility to beam breakup instability (BBU), which is inherent in electron linear accelerators~\cite{Bal1983ICHEA-12,Panofsky_RSI1968}.

It would seem that the plasma could self-consistently focus the witness bunches in the PWFAs, which should improve their output parameters. However, when excited in a overdense plasma, the wave phase region, which admits  simultaneous acceleration and focusing of bunches,  is small~\cite{Ruth:1984pz}, and in the case of underdense plasma, the witness bunch must be placed near the drive bunch, immediately after the bubble~\cite{Rosenzweig_PRA1991}. In addition, focusing of witness positron bunches in the underdense plasma is not possible at all due to the positively charged ion background in the bubble.

To improve the transport of witness bunches, it was proposed to create a vacuum channel in the plasma~\cite{Chiou_PRL1998}, where the transverse defocusing force would be small. However, the hollow-core plasma channel was found to be inherently unstable~\cite{Schroeder1999PRL} because of the BBU instability in case of an off-axis drive bunch. To suppress the instability of witness positron bunches, non-uniform hollow plasma channels were proposed by including in addition a thin coaxial filament~\cite{Pukhov_PRL2018} , or by creating with a laser beam a region of increased plasma density behind the drive bunch~\cite{Reichwein_PRE2022}. Accelerating and focusing forces acting simultaneously on positron bunches can be attained under conditions of a finite plasma column radius, a smaller ion cavity radius (a bubble radius) in a nonlinear regime of wakefield excitation~\cite{Diederichs_PRAB2019}. The focusing fields for positrons in the schemes~\cite{Pukhov_PRL2018,Reichwein_PRE2022,Diederichs_PRAB2019}  are created by forming  behind the bubble the region of increased density of the electrons, which  returned back to the accelerator axis. Other methods for stabilizing the transverse motion of positron bunches have also been proposed (see Section 4C, 4D of the review~\cite{Cao_PRSTAB2024}). Among them, we note the use of bunches with an asymmetric transverse charge density profile~\cite{Zhou_PRL2021}. Under this scheme, the off-axis drive beam-excited quadropole-like wakefield can be dominant in comparison with the dipole-like wakefield.

The proposed schemes for improving the transport of positron bunches are, firstly, complex to practical implementation. Secondly, if we consider them as applied to colliders, different schemes would be needed for the electron and positron arms of the accelerator. Difficulties with positron acceleration forced B. Foster, R. D'Arcy and C. A. Lindstrøm~\cite{Foster_2023} to propose an asymmetric collider design. It is expected that the implementation of such a hybrid collider project HALHF~\cite{Adli:2025hcf} will be much lower in cost than the mature linear-collider designs ILC and CLIC.

This paper proposes and analyzes a modification of the plasma channel method, viz., filling the PWFA bunch transport channel with a plasma of density lower than the density of the outside plasma that creates the accelerating field. This accelerator configuration will be referred to hereafter as a layered plasma wake accelerator (LPWA). It should be noted that a similar plasma waveguide configuration has been previously proposed for the laser beat-wave accelerator scheme, and the accelerator based on this scheme was called a plasma fiber accelerator~\cite{Tajima:1983egt}. In the plasma fiber accelerator configuration considered, the outer plasma density is so high that the electromagnetic wave does not penetrate the outside plasma, and this configuration is effectively a plasma duct for the laser electromagnetic and Langmuir waves. Our proposed configuration of the inhomogeneous plasma structure is different in that we take into account the excitation of eigenwaves in the outside plasma by the drive electron bunch. Our statement of the problem for its application to the PWFA scheme is similar to the one we have already studied for the plasma-dielectric wakefield accelerator (PDWA) --- a dielectric wakefield accelerator with a plasma-filled transport channel~\cite{Sot2014NIMA,Markov2016PAST,Markov_2022,Sot2025NIMA,Berezina2016UFZh}. In the PDWA, the accelerating gradient is created by the excited eigenwave of the dielectric tube, and  the focusing is provided by the Langmuir wave excited in the plasma-filled channel. In this case, for easier and optimal control of the delay time of the witness bunches relative to the drive bunch, it is necessary that the frequencies of the dielectric and Langmuir waves be significantly different. The plasma in the transport channel acts as a passive plasma lens~\cite{Su1990PRA,CHIADRONI2018NIMA,White_2022,Oni2012PAST-PP}, but in the quasi-linear regime of the plasma wave excitation, it also focuses positron bunches~\cite{Markov_2022,GAL2024NIMA,Sot2025NIMA}.

The article is organized as follows. Section~\ref{sec:2} presents the general statement of the problem and gives the parameters of the structure and bunches being in use in the numerical computations of the LPWA. Section~\ref{sec:3} describes the results of analytical studies of the wakefield excitation by a single electron bunch and a regular sequence of equally charged electron bunches. In Section~\ref{sec:4} the results of the 2.5-dimensional PIC-simulation of electron/positron acceleration in LPWA are presented. Section~\ref{sec:5} gives the summary information of the paper.

\section{Statement of the problem}~\label{sec:2}
\begin{figure}[!tbh]
  \centering
  \includegraphics[width=0.48\textwidth]{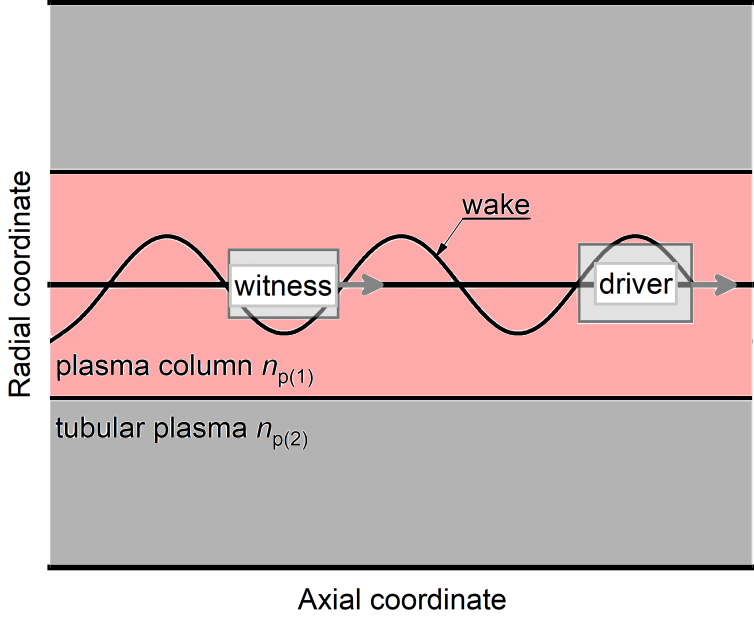}
  \includegraphics[width=0.50\textwidth]{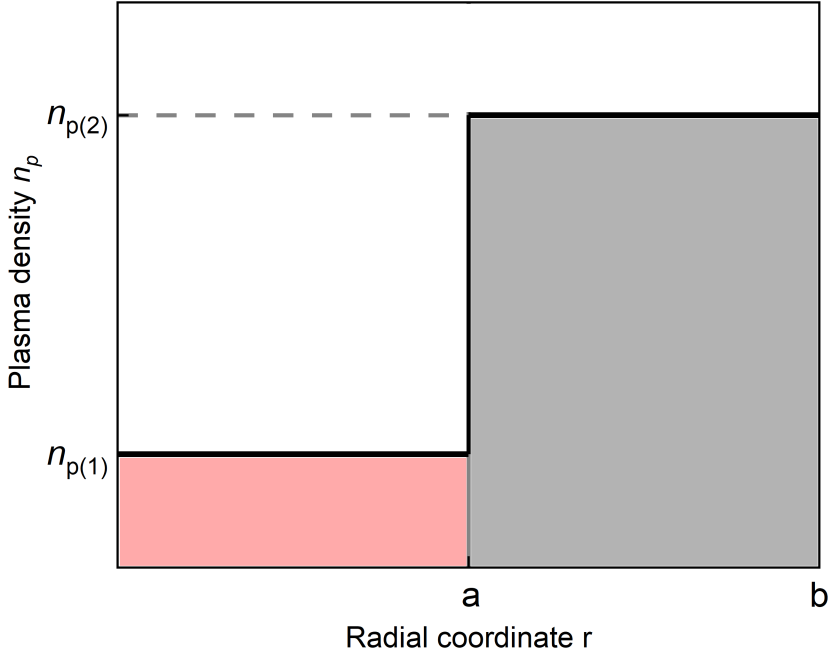}
  \caption{General view of the cylindrical plasma waveguide. Metal coating, layered plasma, drive and witness bunches are shown schematically.  The drive and witness bunches move along to the waveguide axis without an offset.}\label{Fig:01}
\end{figure}

The layered plasma wakefield accelerating structure (LPWA), employed for investigating the wakefield excitation and witness bunch acceleration in it, represents a cylindrical metal waveguide of radius $b$, which is fully filled with a layered cold isotropic plasma (Fig.~\ref{Fig:01}).
The layered plasma comprises two layers adjacent one to another: the inner plasma (the plasma column of radius $a$) has the density $n_{p(1)}$, the density of the outer (tubular) plasma is equal to $n_{p(2)}$. The outer radius of the tubular plasma equals the radius of the waveguide. For ease of carrying out analytical calculations of the excited wakefields and their subsequent comparison with the PIC-simulated data, we assume the interface between the plasma layers to be sharp. The qualitative wakefield distribution would not change if the transition layer thickness is considerably smaller than the length of the excited wave. The waveguide filling with a tubular-shaped dense plasma can be realized with the use of laser beams, which are described by the high-order Bessel functions~\cite{Fan_PRE_2000,Kimura:2011zz,Ges2016Ncom}. No external magnetic field is applied. A drive electron bunch of charge $Q_b$, length $L_b$ and radius $R_b$ is injected into the plasma column. In the case of using a regular sequence of equally charged electron bunches for the wakefield excitation, the total charge of the sequence is equal to the charge of a single bunch.

The witness bunch is injected at some distance from the drive bunches. The goal of the study is to calculate the wakefield excited by the drive electron bunch, to determine the acceptable relationship between the plasma layer densities, which makes possible simultaneous acceleration and focusing of the witness bunches, and eventually, to investigate the transportation of electron/positron bunches in the excited electromagnetic field.

\section{Analytical studies of the wakefield excitation}~\label{sec:3}
We assume that the motion of the on-axis drive electron bunch in the region (1) of the layered plasma ($R_b\le a$) is rectilinear and uniform at a constant velocity $v$. The plasma response to the perturbation by the electron bunch will be described in the quasi-linear approximation (overdense plasma, $n_b \ll n_{p(1),(2)}$, $n_b$ is the electron density of the bunch). The ion dynamics is neglected.

The analytical expressions for the components of the electron bunch-excited electromagnetic field in the layered plasma waveguide (LPW) can be derived by applying the procedure used to derive the expressions for the wakefield in the plasma-dielectric waveguide~\cite{Sot2014NIMA} with the replacement of the dielectric permittivity of the tubular dielectric by the dielectric permittivity of the tubular plasma region (2):
\begin{equation}\label{eq:01}
\begin{split}
\varepsilon(\omega) =
\begin{cases}
  \varepsilon_{p(1)}(\omega) = 1 - \omega_{p(1)}^2/\omega^2,  & \text{$ r \le a$},\\
  \varepsilon_{p(2)}(\omega) = 1 - \omega_{p(2)}^2/\omega^2,  & \text{$a < r \le b$},
\end{cases}
\end{split}
\end{equation}
where $\omega_{p(1,2)} = (4\pi e^2n_{p(1,2)}/m)^{1/2}$ are the respective plasma frequencies of the column and tubular regions of the plasma; $e$ and $m$ are the electron charge and mass, respectively.

Having solved Maxwell's equations with a source in the form of a current density created by a drive bunch, and taking into account the permittivity of the layered medium~(\ref{eq:01}), we obtain the following expressions for the longitudinal component of the electric field $E_z$ and the radial component of the wakefield $W_r = E_r-\beta H_\varphi$ ($E_r$ and $H_\varphi$ are, respectively, the radial electric field and the azimuthal component of the magnetic field, $\beta=v/c$) in the plasma column region (1)
\begin{equation}\label{eq:02}
\begin{split}
E_z(r \leq R_b, \xi ) = &-\frac{4Q_b}{R_bL_b}\left\{ \frac{1}{k_{p(1)}R_b} -\frac{I_0(k_{p(1)}r)}{I_0(k_{p(1)}a)}\Delta_1(k_{p(1)}R_b,k_{p(1)}a) \right\}\Pi_{\parallel}(\xi,\omega_{p(1)}) -\\
&\frac{8Q_b}{R_bL_b}\sum_{s}^{}\frac{v}{\varkappa_{p(1)s}a\omega^2_sD'({\omega_s})}
\frac{I_0(\varkappa_{p(1)s}r)I_1(\varkappa_{p(1)s}R_b)}{I_0^2(\varkappa_{p(1)s}a)}\Pi_{\parallel}(\xi,\omega_s),
\end{split}
\end{equation}

\begin{equation}\label{eq:03}
\begin{split}
E_z(R_b < r \leq a, \xi ) = &-\frac{4Q_b}{R_bL_b} \frac{I_1(k_{p(1)}R_b)}{I_0(k_{p(1)}a)}\Delta_0(k_{p(1)}a,k_{p(1)}r)\Pi_{\parallel}(\xi,\omega_{p(1)}) -\\
&\frac{8Q_b}{R_bL_b}\sum_{s}^{}\frac{v}{\varkappa_{p(1)s}a\omega^2_sD'({\omega_s})}
\frac{I_0(\varkappa_{p(1)s}r)I_1(\varkappa_{p(1)s}R_b)}{I_0^2(\varkappa_{p(1)s}a)}\Pi_{\parallel}(\xi,\omega_s),
\end{split}
\end{equation}

\begin{equation}\label{eq:04}
\begin{split}
W_r(r \leq R_b, \xi ) = &-\frac{4Q_b}{R_bL_b}\frac{I_1(k_{p(1)}r)}{I_0(k_{p(1)}a)}\Delta_1(k_{p(1)}R_b,k_{p(1)}a)\Pi_{\perp}(\xi,\omega_{p(1)}) + \\
&\frac{8Q_b}{R_bL_b}\sum_{s}^{}\frac{v^2}{a\omega^3_sD'({\omega_s})}
\frac{I_1(\varkappa_{p(1)s}r)I_1(\varkappa_{p(1)s}R_b)}{I_0^2(\varkappa_{p(1)s}a)}\Pi_{\perp}(\xi,\omega_s),
\end{split}
\end{equation}

\begin{equation}\label{eq:05}
\begin{split}
W_r(R_b < r \leq a, \xi ) = &-\frac{4Q_b}{R_bL_b}\frac{I_1(k_{p(1)}R_b)}{I_0(k_{p(1)}a)}\Delta_1(k_{p(1)}r,k_{p(1)}a)\Pi_{\perp}(\xi,\omega_{p(1)}) + \\
&\frac{8Q_b}{R_bL_b}\sum_{s}^{}\frac{v^2}{a\omega^3_sD'({\omega_s})}
\frac{I_1(\varkappa_{p(1)s}r)I_1(\varkappa_{p(1)s}R_b)}{I_0^2(\varkappa_{p(1)s}a)}\Pi_{\perp}(\xi,\omega_s),
\end{split}
\end{equation}
and also in the tubular plasma region (2):
\begin{equation}\label{eq:06}
\begin{split}
E_z(a < r \leq b, \xi ) = \frac{8Q_b}{R_bL_b}\sum_{s}^{}\frac{v}{\varkappa_{p(1)s}a\omega^2_sD'({\omega_s})}
\frac{I_1(\varkappa_{p(1)s}R_b)}{I_0(\varkappa_{p(1)s}a)}\frac{\Delta_0(\varkappa_{p(2)s}r,\varkappa_{p(2)s}b)}{\Delta_0(\varkappa_{p(2)s}a,\varkappa_{p(2)s}b)}\Pi_{\parallel}(\xi,\omega_s),
\end{split}
\end{equation}

\begin{equation}\label{eq:07}
\begin{split}
W_r(a < r \leq b, \xi ) = \frac{8Q_b}{R_bL_b}\sum_{s}^{}\frac{(1-\beta^2\varepsilon_{p(2)s})}{a\varkappa_{p(1)s}\varkappa_{p(2)s}\omega_sD'({\omega_s})}
\frac{I_1(\varkappa_{p(1)s}R_b)}{I_0(\varkappa_{p(1)s}a)}\frac{\Delta_1(\varkappa_{p(2)s}r,\varkappa_{p(2)s}b)}{\Delta_0(\varkappa_{p(2)s}a,\varkappa_{p(2)s}b)}\Pi_{\perp}(\xi,\omega_s),
\end{split}
\end{equation}

The time and the longitudinal coordinate dependence of the excited wakefield components is described by the functions $\Pi_{\parallel}(\xi,\omega)$ and $\Pi_{\perp}(\xi,\omega)$, which take the form
\begin{equation}\label{eq:08}
\begin{split}
\Pi_{\parallel}(\xi,\omega) = &\theta (\xi)\sin\omega\xi - \theta (\xi  - T_b)\sin\omega(\xi - T_b),\\
\Pi_{\perp}(\xi,\omega) = &\theta (\xi)(1 - \cos\omega\xi) - \theta (\xi - T_b)(1 - \cos \omega(\xi - T_b)),
\end{split}
\end{equation}
where $\xi =t-z/v$, $t$ is the time, $z$ is the longitudinal coordinate, $T_b=L_b/v$ is the bunch duration,  $\theta(x)$ is the Heaviside step function.

The function $D'(\omega_s)=d D(\omega_s)/d\omega_s$ involved in the denominators of the expressions~(\ref{eq:02})-(\ref{eq:07}) is the derivative of the dispersion function calculated at the Cherenkov resonance points $k_z=\omega/v$ of the beam with the eigenwaves of the layered plasma waveguide. The mentioned eigenwaves are determined from the solution of the dispersion equation
\begin{equation}\label{eq:09}
\begin{split}
D(\omega,k_z) \equiv \frac{\varepsilon_{p(1)}}{\varkappa_{p(1)}}\frac{I_1(\varkappa_{p(1)}a)}{I_0(\varkappa_{p(1)}a)} - \frac{\varepsilon_{p(2)}}{\varkappa_{p(2)}}\frac{\Delta_1(\varkappa_{p(2)}a,\varkappa_{p(2)}b)}{\Delta_0(\varkappa_{p(2)}a,
\varkappa_{p(2)}b)} =  0,
\end{split}
\end{equation}
where $\varkappa_{p(1)} = k_z^2 - \omega^2/c^2\varepsilon_{p(1)}(\omega)$,\, $\varkappa_{p(2)} = k_z^2 - \omega^2/c^2\varepsilon_{p(2)}(\omega)$.
The rest notation in~(\ref{eq:02})-(\ref{eq:07}) $k_{p(1)}=\omega_{p(1)}/v$,\, $\varkappa_{p(1)s} = \varkappa_{p(1)}(k_z=\omega/v,\omega=\omega_s)$, \, $\varkappa_{p(2)s} = \varkappa_{p(2)}(k_z=\omega/v,\omega=\omega_s)$,\, $ \Delta_0(x,y) = I_0(x)K_0(y) - K_0(x)I_0(y)$,\, $\Delta _1(x,y) = I_1(x)K_0(y) + K_1(x)I_0(y)$, and $I_{0,1}$, $K_{0,1}$ are the modified Bessel and MacDonald functions of the $0^{th}$ and $1^{st}$ orders, respectively. In the derivation of the expressions~(\ref{eq:02})-(\ref{eq:08}) it was assumed that the electron bunch had a rectangular electron density distribution profile in the longitudinal $\xi$ and transverse $r$ coordinates.

The expressions~(\ref{eq:02})-(\ref{eq:05}) demonstrate the fact that the total wakefield in the plasma column region comprises the wakefield of the Langmuir wave at the frequency $f_{p(1)}=\omega_{p(1)}/2\pi$ and the wakefield of the modes resonant with the drive bunch at frequencies $\omega_{s}$ ($TM$-mode). In the tubular plasma region, the wakefield comprises only the modes resonant with the drive bunch at frequencies $\omega_{s}$, because the drive bunch does not penetrate this area and, as a consequence, does not excite the Langmuir wave at the frequency $\omega=\omega_{p(2)}$.

\begin{figure}[!tbh]
  \centering
  \includegraphics[width=0.5\textwidth]{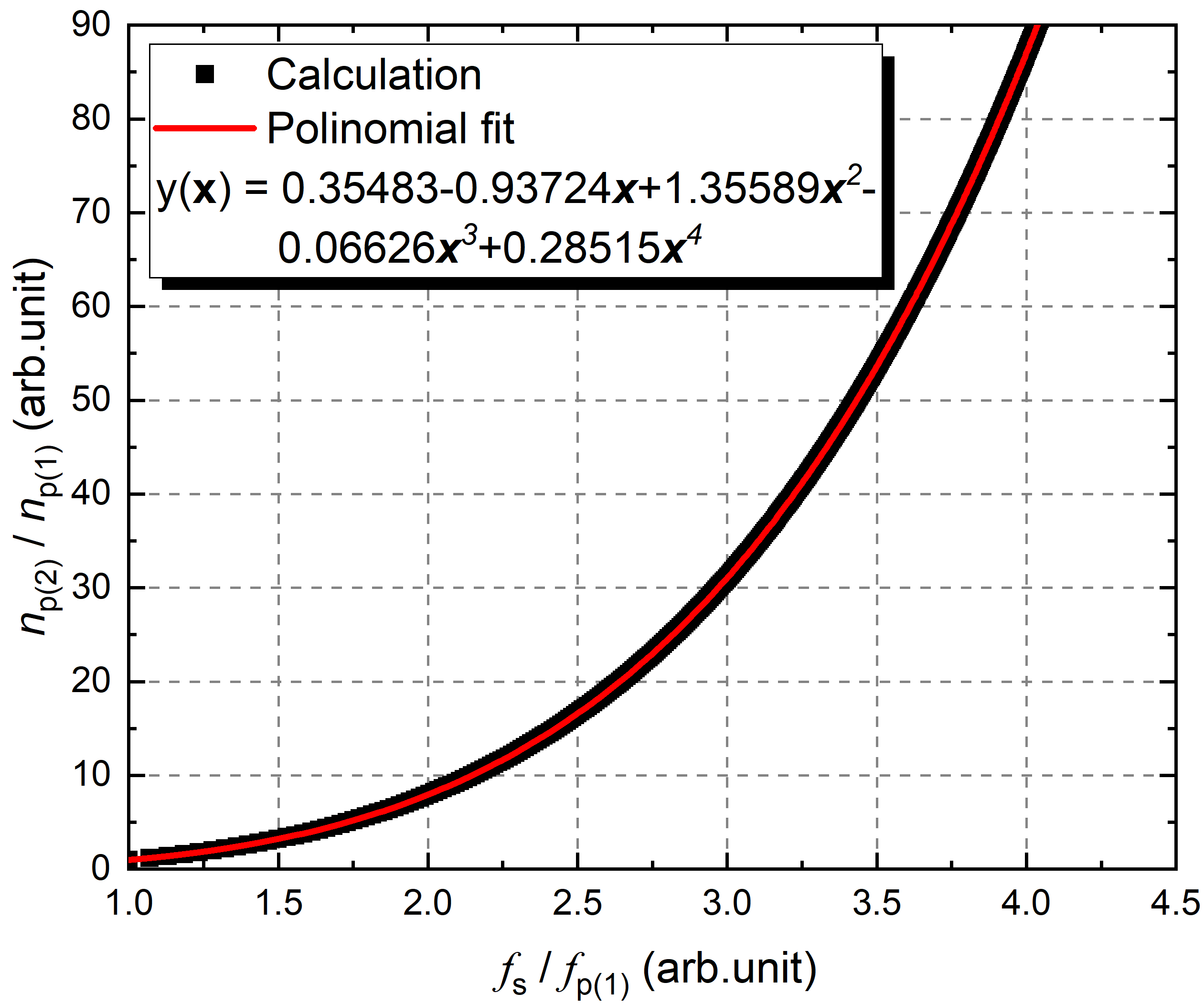}
  \caption{Tubular plasma density versus resonant eigenfrequency of the layered plasma waveguide. The waveguide dimensions are given in Tab.~\ref{tab:01}. Plasma density in the plasma column
region (1) is $n_{p(1)}=4.93\cdot 10^{14}\rm cm^{-3}$.}\label{Fig:02}
\end{figure}

We now turn to the numerical analysis of the derived expressions for the electromagnetic field components, excited by the drive electron bunch.
The parameters of the LPW and the drive bunch are given in ~Tab.\ref{tab:01}. Here we have chosen the plasma column density to be $4.93\cdot 10^{14}\rm cm^{-3}$, this corresponding to the plasma frequency equal to 199.3 GHz, or the Langmuir wave length $1.5$ mm. By varying the tubular plasma density and, in accord with the dispersion equation~(\ref{eq:09}) and the Cherenkov resonance condition  $k_z=\omega/v=2\pi f/v$,  we shall change the resonant frequencies of the layered plasma waveguide. As numerical calculations demonstrate, the dispersion equation~(\ref{eq:09}) has the only one solution, that satisfies the condition of the Cherenkov resonance. Figure~\ref{Fig:02} illustrates the dependence of the resonant frequency of the layered plasma waveguide on the coaxial plasma density $n_{p(2)}$. The resonant TM-mode frequency shows a polynomial dependence. As noted above, for easier control over the processes of witness-bunch acceleration and focusing, the frequencies ratio $f_{s}/f_{p(1)}$ must be high. We choose it to be equal to three. Then, as it follows from Fig.~\ref{Fig:02}, the tubular plasma density is equal to $n_{p(2)}=1.53\cdot 10^{16}\rm cm^{-3}$. This value of the tubular plasma density will be used in our further calculations.

\begin{figure}[!tbh]
  \centering
  \includegraphics[width=0.5\textwidth]{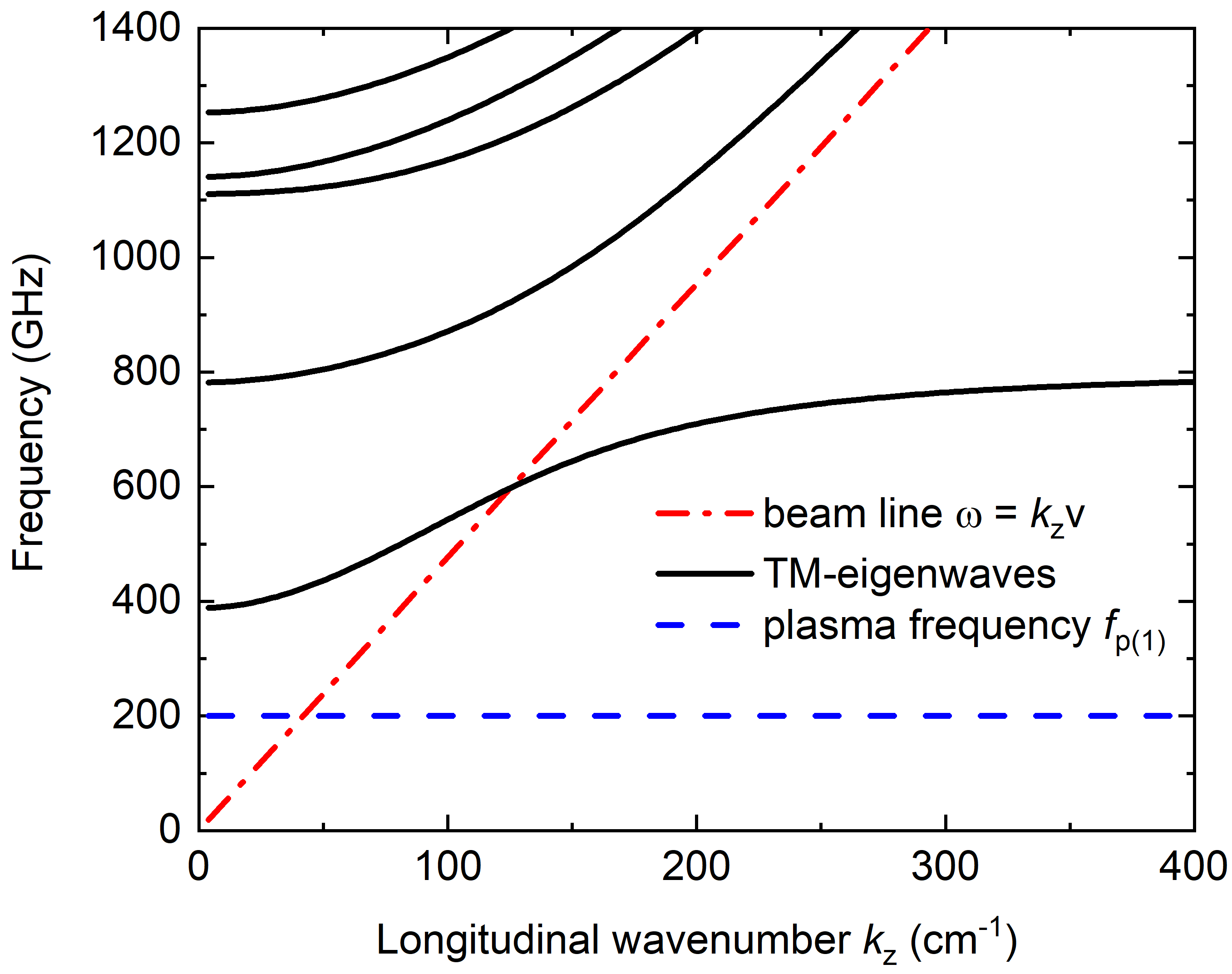}
  \caption{Dispersion dependence of the layered plasma waveguide. Parameters of the waveguide and the plasma density in the plasma column region $n_{p(1)}$  are presented in Tab.~\ref{tab:01}. The tubular plasma density (region 2) is $n_{p(2)}=1.53\cdot 10^{16}\rm cm^{-3}$. The dash-dotted red line is the Cherenkov line $2\pi f=k_zv_0$, the dashed blue line $f=f_{p(1)}$ shows the Langmuir oscillations. Among the multitude of electromagnetic $TM_{0n}$-waves (solid black lines) there exists only one slow $TM_{01}$ wave, resonant with the relativistic electron bunch ($W_b=5\,GeV$). The bunch resonance with the Langmuir wave $\omega_{p(1)}=k_zv_0$ describes the quasistatic waves responsible for focusing the accelerated bunches (the first summands in the expressions~(\ref{eq:02})-(\ref{eq:05})).}\label{Fig:03}
\end{figure}

\begin{table}
\caption{Parameters of the layered plasma waveguide and bunches used in the numerical calculations}
\centering
\begin{tabular}{l c c}
\hline
Parameter & Value &Unit \\
\hline
Waveguide radius $b$                & $600$                & $\rm \mu m$     \\
Plasma column radius $a$            & $300$                & $\rm \mu m$     \\
Plasma column density $n_{p(1)}$    & $4.93\cdot 10^{14}$  & $\rm cm^{-3}$   \\
Energy of bunch $W_b$               & $5$                  & $\rm GeV$       \\
Charge of bunch $Q_b$               & $2$                  & $\rm nC$        \\
Length of bunch $L_b$               & $250$                & $\rm \mu m$     \\
Radius of bunch $R_b$               & $250$                & $\rm \mu m$     \\
\hline
\end{tabular}
\label{tab:01}
\end{table}
The general view of the dispersion dependencies $f(k_z)$ of the layered plasma waveguide at the found $n_{p(2)}$ value is given in Fig.~\ref{Fig:03}.
Figure~\ref{Fig:03} clearly demonstrates the preceding, that the beam mode crosses only one dispersion curve of the $TM_{0n}$ waves.
The resonant (Cherenkov) frequency $f_{1}$ for the given set of the parameters equals $f_{1} = 598\,\mathrm{GHz}$. The other waves are defined as fast.
It should be noted that the waveguide under discussion has only one resonant $TM$-mode at any thickness of the tubular plasma and arbitrary densities of the plasma column and the tubular plasma. Therein lies the essential difference from the plasma-dielectric waveguide~\cite{Sot2014NIMA,Markov_2022,Sot2025NIMA}, where a single-mode excitation of  $TM$ waves is realized only for a thin dielectric.
In the case of a thick dielectric, aside from the fundamental accelerating mode $TM_{01}$, the excitation of higher radial $TM$-modes takes place.

\begin{figure}[!th]
  \centering
  \includegraphics[width=0.49\textwidth]{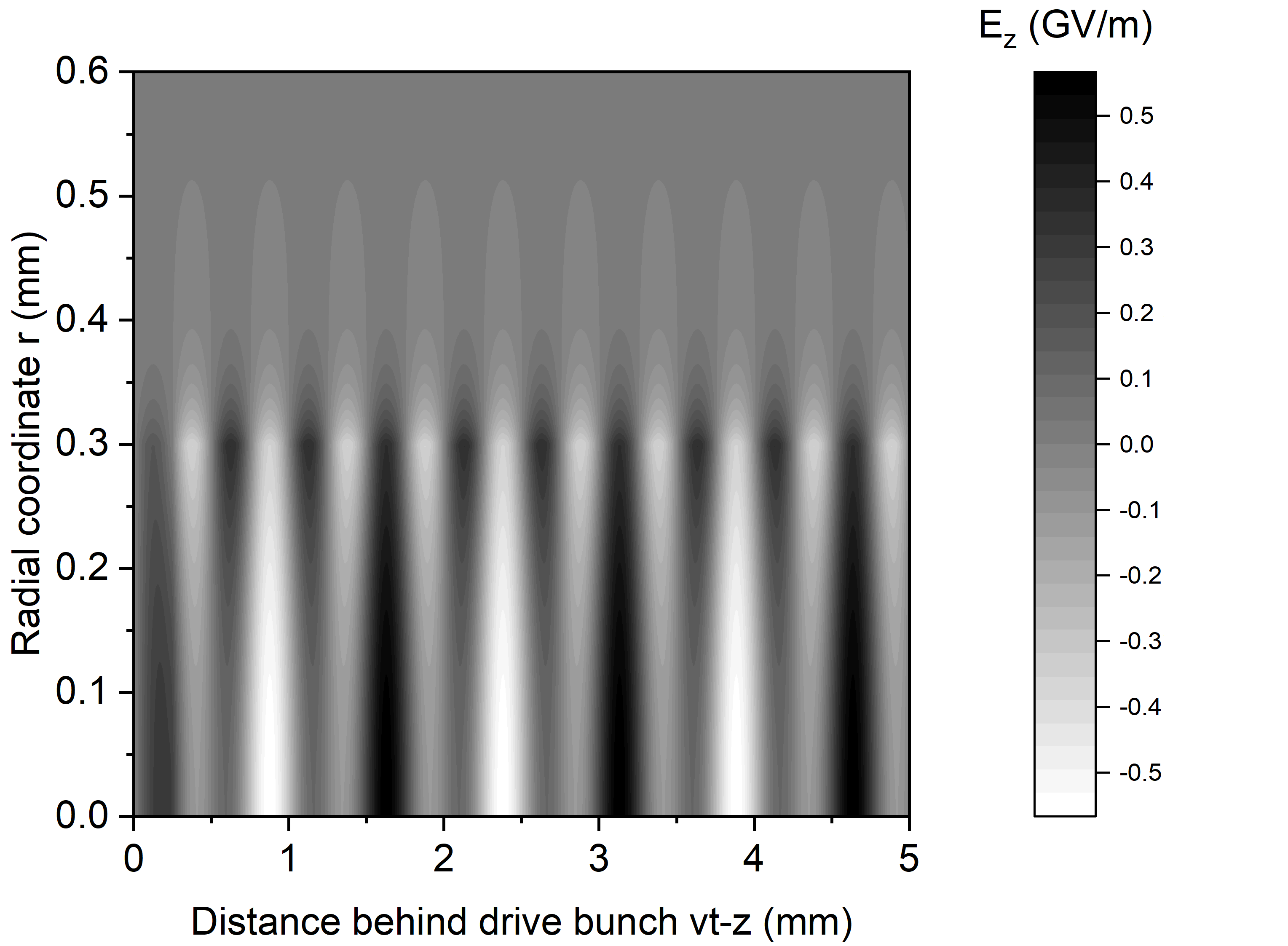}
  \includegraphics[width=0.49\textwidth]{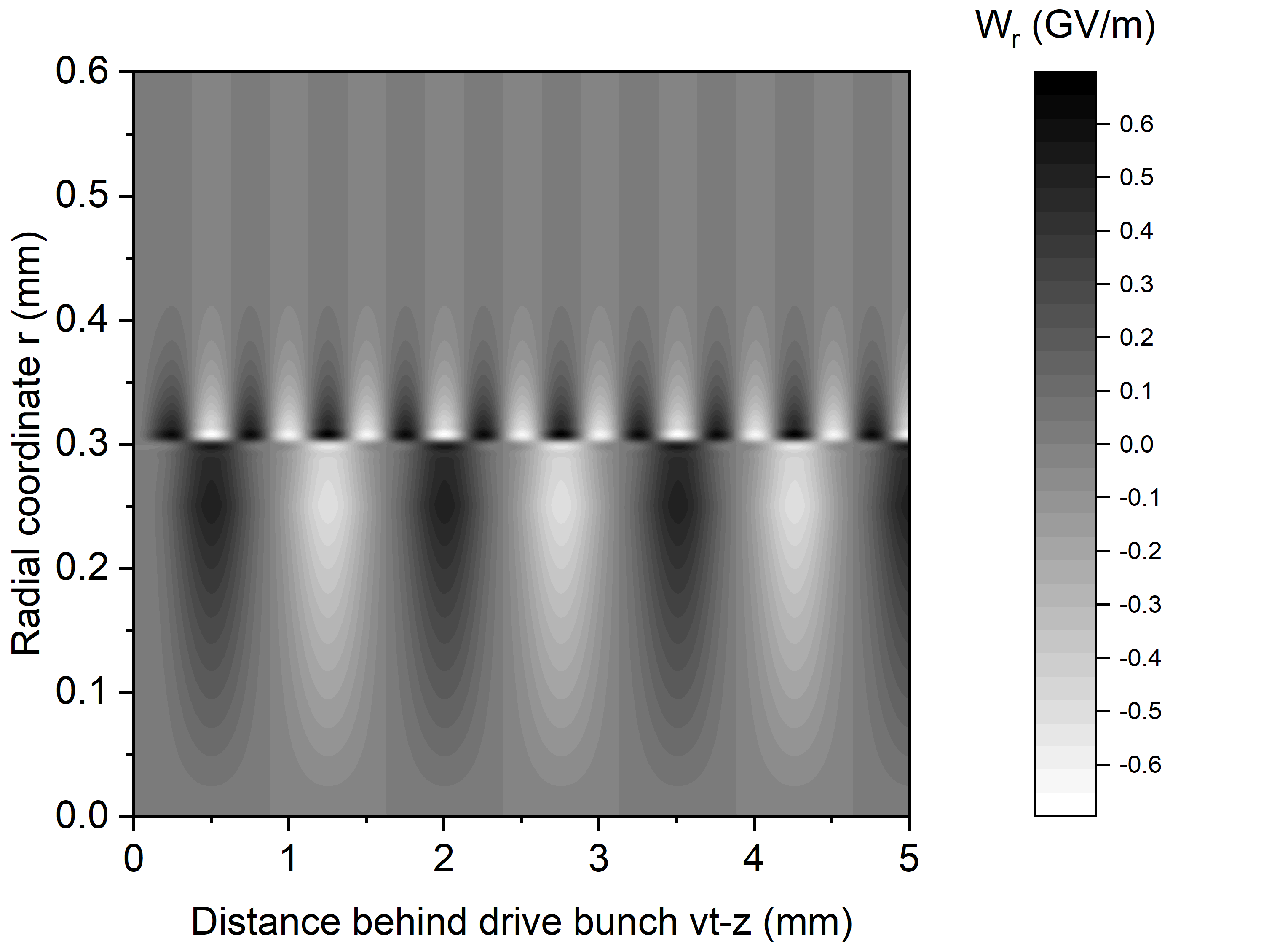}
  \caption{Two-dimensional distributions of the longitudinal $E_z$ and traverse $W_r$ wakefields excited by a single drive electron bunch. Drive bunch head is located at $\xi = vt - z = 0$.  In the plasma column
region (1) ($r<a=0.3\,mm$) the period of the transverse wakefield  $W_r$, focusing the drive and witness bunches is equal to the Langmuir wave period $\approx 1.5\,\rm mm$. In the same region, the basic period of the longitudinal wakefield $E_z$, that accelerates the witness bunches, is equal to the period of the $TM_{01}$ wave $\approx 0.5\,\rm mm$.}\label{Fig:04}
\end{figure}

We now go directly to the analysis of the excited wakefield. Figure~\ref{Fig:04} shows the two-dimensional distributions of axial $E_z$ and radial $W_{r}$ wakefields excited by a single drive electron bunch. It can be seen that the most essential qualitative differences between the longitudinal and transverse wakefields distributions, namely, in their longitudinal structure, are observed in the region of the plasma column (region 1) of the layered plasma waveguide. In this region, the transverse wakefield $W_r$ has a clearly defined longitudinal period equal to $\approx 1.5\,\rm mm$, which corresponds to the Langmuir wave frequency $\approx 200\,\rm GHz$. In the distribution of the longitudinal wakefield $E_z$ in the region (1) two periods can be seen, viz., $\approx 1.5\,\rm mm$ and $\approx 0.5\,\rm mm$. The second period is due to the excitation of the $TM_{01}$ wave of the layered plasma waveguide, having the frequency $\approx 600\,\rm GHz$ (see Fig.~\ref{Fig:03}). In the tubular plasma region, the distributions of the longitudinal and transverse wakefields have a similar longitudinal structure with the period equal to the period of the $TM_{01}$ wave. Their amplitudes decline rapidly from the interface inward the tubular plasma in consequence of the surface character of the $TM_{01}$ wave.

\begin{figure}[!th]
  \centering
  \includegraphics[width=0.5\textwidth]{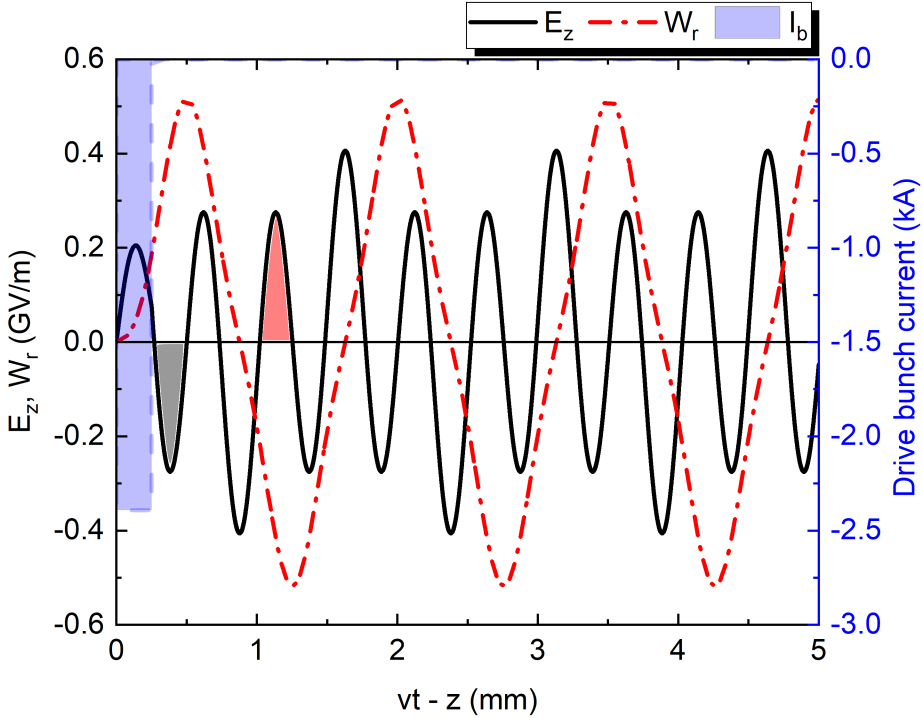}
  \caption{Axial structure of the longitudinal (black solid line) and radial (dash-dot red line) wakefields behind the single drive electron bunch with a charge of $2\,\rm nC$. The fill areas show the position and current of the drive bunch (blue ground-color) and possible positions for accelerating the electron (grey ground-color) and positron (pink ground-color) witness bunches. The witness bunches centers are found in the peak of the accelerating field.}\label{Fig:05}
\end{figure}

To demonstrate the longitudinal structure of the excited field more clearly, Fig.~\ref{Fig:05} shows the distribution of axial and radial wakefields in the region of the plasma column on the surface of the drive bunch $r = R_b$.
It is obvious from the figure that the distribution of the transverse wakefield is close to monochromatic in character, whereas the longitudinal profile of the axial wakefield has the character of low-frequency modulation with a period three times greater than the period of the resonant mode. It should be noted that at the chosen parameters of the drive bunch and the plasma waveguide at the points of maximum of the longitudinal electric field, the transverse wakefield $W_r$ vanishes. However, the substantial difference between spatial periods of longitudinal and transverse wakefield amplitudes makes it possible to find behind the drive bunch the regions, wherein the test bunches (both electron and positron) could be located for their simultaneous acceleration and focusing. The maximum of the acceleration field on the surface of the drive bunch envelope equals $\approx 300\,\mathrm{MV/m}$.

\begin{figure}[!tbh]
  \centering
  \includegraphics[width=0.49\textwidth]{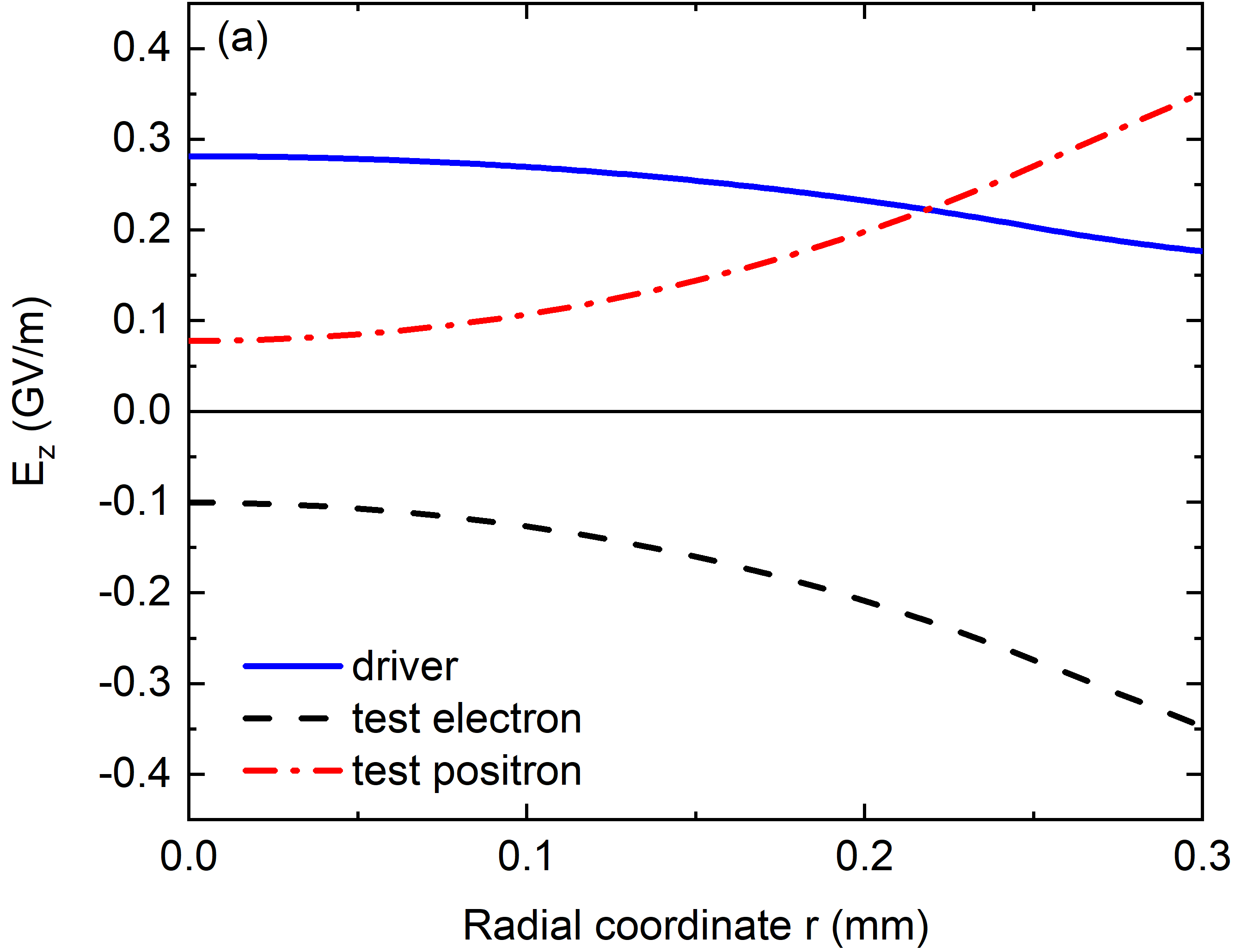}
  \includegraphics[width=0.49\textwidth]{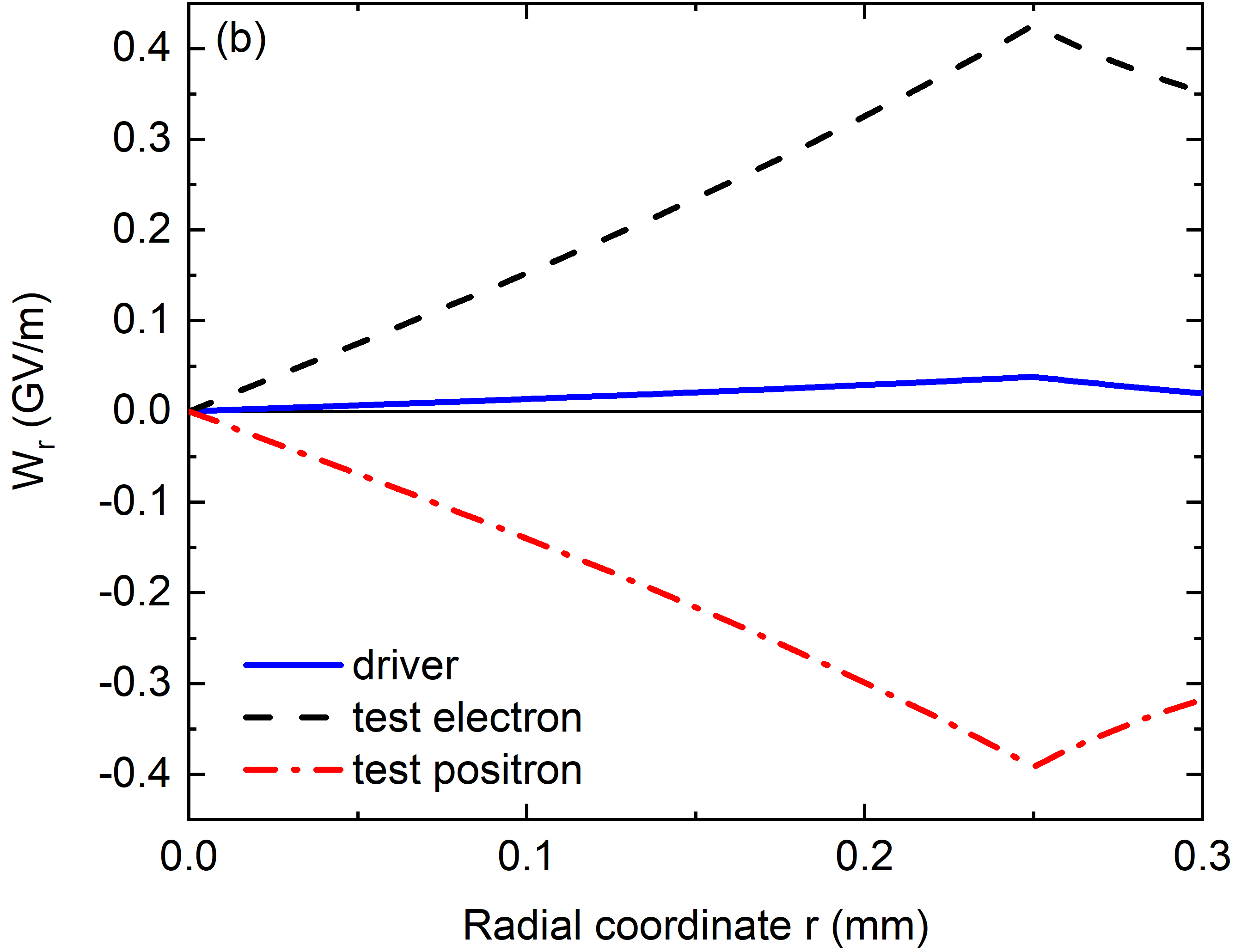}
  \caption{Transverse distribution of the axial (a) and radial (b) wakefields in the plasma column region for three values of the axial distance behind the drive bunch head, namely: the centre of the driver ($vt-z = L_b/2$), the maximum of the accelerating field for the test electron ($\xi=0.39\,\rm mm$) and positron ($\xi=1.12\,\rm mm$) bunches. Blue solid lines depict the wakefield acting on the drive bunch, black dash-dotted lines depict the wakefield acting on the electron witness bunch and red dashed lines depict the wakefield acting on the positron witness bunch. The transverse wakefield is linearly dependent on the radial coordinate.}\label{Fig:06}
\end{figure}

Figure~\ref{Fig:06} shows the transverse distribution of the axial and radial wakefields inside the plasma column, region (1) of the layered plasma waveguide.
For the drive bunch, the maximum value of the axial wakefield  corresponds to the waveguide axis, decreasing to the periphery of the bunch transportation channel, region (1). By contrast, for the witness electron and positron bunches, the longitudinal wakefield increases from the axis towards the interface of the two regions of the layered plasma. The distinctive behavior of the $E_z$ dependence on the radial coordinate in the positions of the drive and witness bunches is due to different contributions of the Langmuir wave and the surface $TM_{01}$ wave to the total field.
At the drive bunch position, the dominant contribution comes from the Langmuir wave decreasing from the waveguide axis (the first summand in the equation~(\ref{eq:02})), whereas at the position of the witness bunches the predominance belongs to the wave decreasing from the boundary surface of the two plasmas. The transverse wakefield for the drive bunch is equal to zero on the waveguide axis, and then increases by the linear law, reaching its maximum value on the drive bunch envelope. The same relationship is also observed for the transverse wakefield at possible locations of accelerated electron and positron witness bunches (see Fig.~\ref{Fig:05}). If the radius of witness bunches is chosen to be equal to the radius of the drive bunch, then the focusing field on their surfaces will be equal to $\approx \rm 400\,MV/m$.

\begin{figure}[!th]
  \centering
  \includegraphics[width=0.49\textwidth]{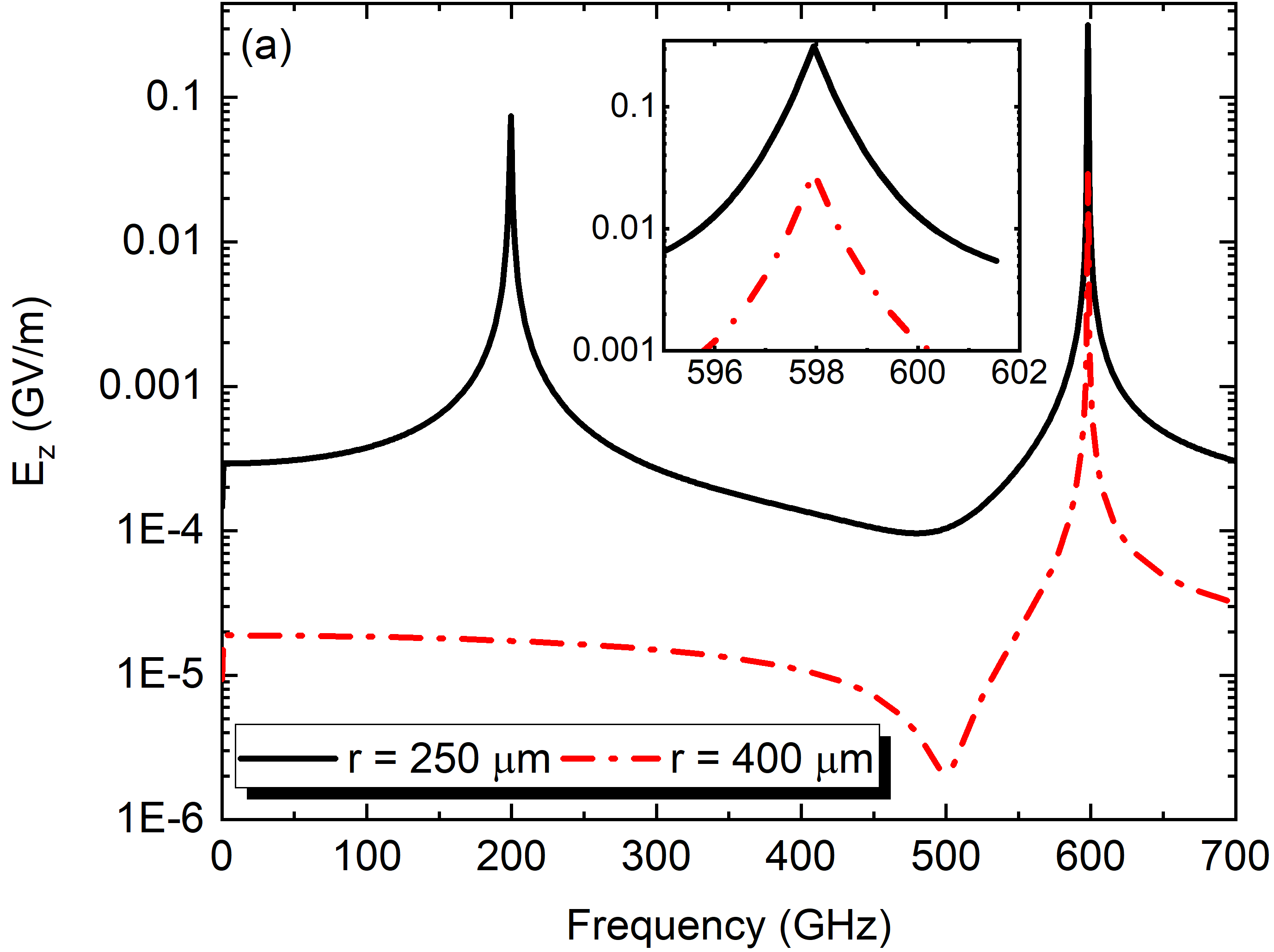}
  \includegraphics[width=0.49\textwidth]{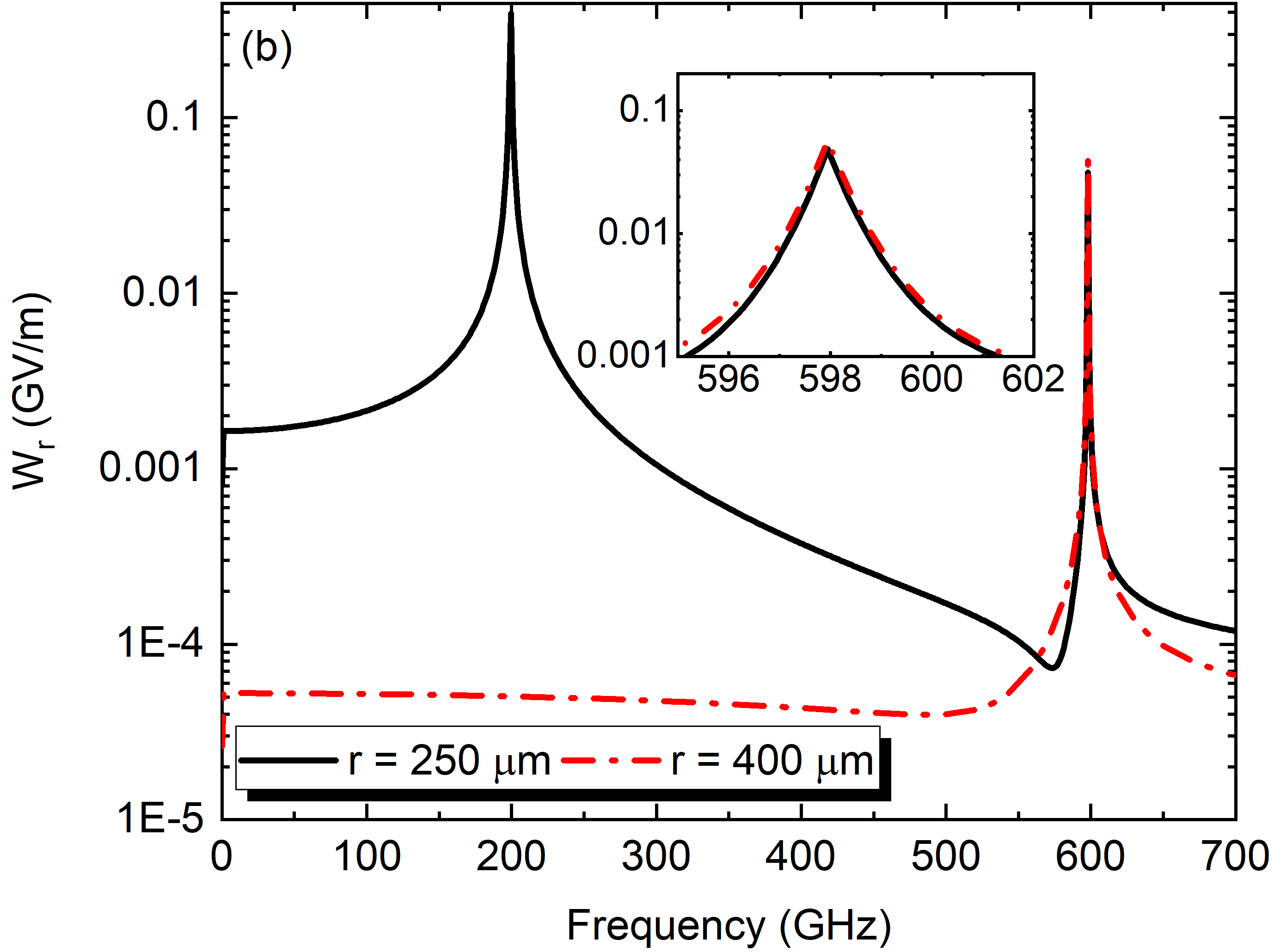}
  \caption{Amplitude-frequency spectra of the longitudinal (a) and transverse (b) wakefields excited by a single drive electron bunch calculated in the plasma column region at $r = 250\, \rm \mu m$ (solid black line) and in the tubular plasma region at $r = 450\, \rm \mu m$ (dash-dotted red line). In the particles transportation channel ($r<a$), the main contribution to the transverse wakefield comes from the Langmuir wave at the frequency $f=f_{p(1)}$, whereas the longitudinal field is mainly contributed by the resonant wave of the layered plasma waveguide.}\label{Fig:07}
\end{figure}

Figure~\ref{Fig:07} shows the spectra of the excited wakefield.  The spectra of the longitudinal $E_z$ and transverse $W_r$ components of the excited wakefield have been calculated in the plasma column region ($r = 250\, \rm \mu m$) and in the tubular plasma region ($r = 450\, \rm \mu m$).
In the plasma column region, the longitudinal and transverse wakefield spectra demonstrate two maxima, which correspond to the Langmuir frequency of the plasma column $f_{p(1)}$ and the resonant eigenfrequency of the waveguide $f_{1}$. The frequency that gives the main contribution to the longitudinal wakefield is the resonant frequency $f_{1}$. The spectral amplitude at the Langmuir frequency $f_{p(1)}$ is by a factor of three lower than the spectral amplitude at the resonant frequency $f_{1}$, and the excitation of this wave results in the modulation of the amplitude profile of the longitudinal wakefield. The radial wakefield spectrum shows that in the plasma column region the main contribution corresponds to the Langmuir frequency $f_{p(1)}$, and the contribution to its total amplitude at the resonant frequency $f_{1}$ is negligibly small. In the tubular plasma region, the spectra of both the longitudinal and transverse wakefields have a single maximum corresponding to the resonant frequency $f_{1}$. The maximum in the wakefield spectrum at the Langmuir frequency of the tubular plasma $f_{p(2)}=1.1\, \rm THz$ is absent, because the drive bunch is localized outside this region, and no excitation takes place at this frequency.

One of the ways of generation high-amplitude wakefields is to use a regular sequence of bunches injected so that the fields from individual bunches add up coherently~\cite{Chen_PRL1985,Onishchenko1995782,Sheinman_2008JTF,ONISH_2016NIMA,Verra_2025PRE}. For this purpose, the bunches are injected at the resonance frequency, i.e., the spacing between separate bunches in their sequence is equal to the resonant wavelength. As an illustration, let us consider the sequence of four electron bunches with a charge of $0.5\, \rm nC$ each, the bunch spacing in the sequence being $\lambda_1=v/f_1$. To derive the expression for the wakefield of this regular sequence of bunches, the functions $\Pi_{\parallel}(\xi,\omega)$ and $\Pi_{\perp}(\xi,\omega)$ in the expressions~(\ref{eq:02})-(\ref{eq:07}) should be replaced by:
 \begin{equation}\label{eq:21}
\begin{split}
\Pi_{\parallel}^i(\xi_i,\omega) = & \Pi_{\parallel}(\xi - (i-1)T_r,\omega),\\
\Pi_{\perp}^i(\xi_i,\omega) = & \Pi_{\perp}(\xi - (i-1)T_r,\omega),
\end{split}
\end{equation}
where $T_r = 1/f_1$, with a further summation over all the bunches $i=1\ldots4$, $i$ is the bunch number.

\begin{figure}[!tbh]
  \centering
  \includegraphics[width=0.49\textwidth]{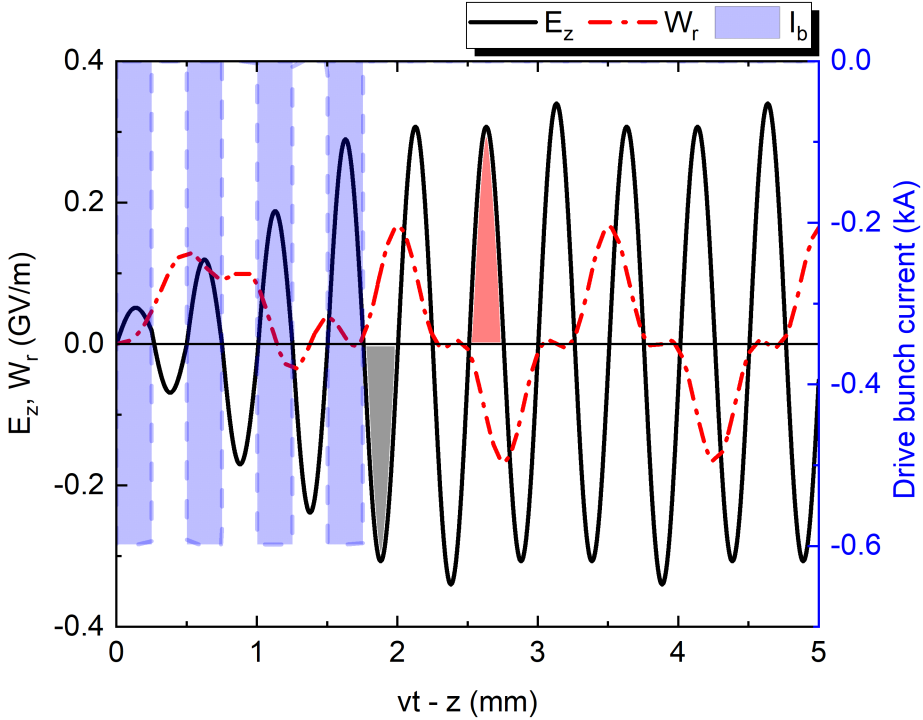}
  \caption{Axial structure of the longitudinal and radial wakefields excited by the resonant sequence of four drive electron bunches with an overall charge of $2\,\rm nC$. The bunch positions and current are depicted by blue rectangles. The black solid line shows the longitudinal electric field $E_z$, the red dash-dotted line displays the radial wakefield $W_r$. The amplitude of the longitudinal electric field behind the last bunch of the sequence is approximately the same as that for the case of excitation of the layered plasma waveguide by a single bunch (see Fig.~\ref{Fig:05}).}\label{Fig:08}
\end{figure}

Figure~\ref{Fig:08} shows the longitudinal amplitude distribution of axial and radial wakefields excited by the resonant sequence of four drive electron bunches with a charge of $0.5\,\rm nC$ each (overall charge is $2\,\rm nC$).
The longitudinal wakefield amplitude grows linearly from the first bunch to the last one in the sequence.  The amplitude increases due to coherent summation of the wakefields excited individually by each bunch, since the main contribution to the longitudinal wakefield amplitude comes from  the resonant frequency $f_1$ (the bunch injection frequency). After the last bunch of the sequence, the amplitude profile of the axial wakefield appears practically regular with insignificant modulation. The reason of this weak amplitude modulation lies in the Langmuir wave excitation, which also contributes to the total excited wakefield. As expected, the amplitude of the longitudinal wakefield is practically coincident with the amplitude of the wakefield generated by the single drive bunch (compare with Fig.~\ref{Fig:05}), i.e., the regular bunch sequence can provide the same acceleration rate, as has the single bunch of the same charge. The amplitude of the radial wakefield does not increase, because the main contribution to this component is brought not by the resonant mode of frequency   $f_{1}$ (at which the sequence bunches are injected) but by the Langmuir wave of frequency $f_{p(1)}$.
From the fact that for the chosen parameters of the layered plasma waveguide (see~Tab.\ref{tab:01}) and the tubular plasma density $n_{p(2)}=1.53\cdot 10^{16}\rm cm^{-3}$, the Langmuir wavelength of the plasma column is $\lambda_{p(1)}=3\lambda_1$, the radial focusing of drive bunches is non-uniform (inhomogeneous) along their sequence. The analysis has demonstrated that the focusing uniformity of regular sequence bunches can be improved through increasing the Langmuir wavelength of the plasma column.

Thus the linear theory of wakefield excitation predicts that in the layered plasma waveguide, in both cases, either it is excited by a single drive electron bunch, or by a regular sequence of equally charged drive electron bunches, simultaneous acceleration and focusing of both electron and positron witness bunches can be realized.

\section{PIC-simulation of the LPWA}~\label{sec:4}
To simulate numerically the wakefield excitation and the dynamics of bunches, both drivers and witnesses, we have used our own 2.5-dimensional code for particles in cells (PIC)~\cite{Sot2014NIMA,Sot2025NIMA}. Cell size  $hr \times hz = 1.432\,\mu \rm m \times 3.129\,\mu \rm m$. As boundary conditions (BC) at the input and output ends of the LPWA, we have chosen the conditions of setting the tangential electric field components to zero on the cylindrical surface, input and output parts of the waveguide  (electric BC). The diameter of the cylindrical metal waveguide, as well as the diameters and densities of the plasma column and tubular plasma regions of the layered plasma were chosen to be the same as the ones in the analytical calculations of the excited wakefield given in the preceding section.  The injection of electron driver bunches into the plasma column region (1) was followed by the injection of accelerated electron or positron bunch (witness bunch) into the excited wakefield. Similarly to the analytical calculations in section~\ref{sec:3}, we have performed PIC simulation for a single drive bunch with the charge $-2\,\textrm{nC}$, and a regular sequence consisting of four drive bunches, each with the charge $-0.5\,\textrm{nC}$. The injection repetition rate of the drive bunches $T_r$ is equal to the time period of the wakefield of the resonant mode $TM_{01}$  of the layered waveguide, defined in the previous section, $T_r=1/f_1$. After the drive bunches, with some delay, the electron or positron witness-bunch with the charge $\mp 0.2\,\textrm{nC}$ was injected into the plasma column region (1).  The diameters of the injected drivers and the witness were the same in value and were equal to  $501.2\,\rm \mu m$. The drive bunch length was one-half the resonance wave length ($\lambda_1=250.7\,\rm \mu m$), while the witness length was by a factor of two shorter ($125.9\, \rm \mu m$). The transverse charge distribution in the bunches was set to be uniform, and the longitudinal charge distribution in the bunches -- to be Gaussian. At that, the FWHM drive bunch size was equal to $125\,\rm \mu m$, while the witness-bunch size was equal to $62.7\,\rm \mu m$. The delay times of the electron $t_{del}^{e^-}$ and positron $t_{del}^{e^+}$ witness bunches relative to the start of drive bunch injection were calculated in the same way as in~\cite{Markov_2022} in order to attain the highest gain in the witness-bunch energy. The time delay values of the witness-bunches for the cases of a single drive bunch and a sequence of drive bunches are given in Table~\ref{Tabl_2}. The length of the waveguide section in the case of using a sequence of bunches as a drive was different from that in the case of a single drive bunch. This was due to the necessity of providing approximately the same distances of witness-bunch acceleration in order to make correct comparison between the energy gains attained.
\begin{table}[!th]
   \centering
   \caption{Types of driver and delay times of witness bunch injection relative to the start of drive sequence injection}
   \begin{tabular}{c|l|l|p{0.7in}}
   \hline
   \hline
       \textbf{Case} &
       \begin{tabular}[t]{p{1.8in}c}
         \textbf{Drive sequence}
       \end{tabular}
       &
       \begin{tabular}[t]{p{0.7in}|l}
                          \textbf{Witness} & \textbf{Delay time, ps}\\
                          &
       \end{tabular}
       &
       \textbf{Length of LPW, mm}
       \\
   \hline
   \hline
       1 &
       \begin{tabular}[t]{p{1.8in}}
         single drive bunch with a charge equal to $-2\,\textrm{nC}$
       \end{tabular}
       &
       \begin{tabular}[t]{p{0.7in}|l}
                         electron & 1.17 \\
                         positron & 3.63 \\
       \end{tabular}
       &
       16
       \\
   \hline
       2 &
       \begin{tabular}[t]{p{1.8in}}
          4 drive bunches, each with a charge of $-0.5\,\textrm{nC}$, repetition rate is 1.672~ps
       \end{tabular}
       &
       \begin{tabular}[t]{p{0.7in}|l}
                      electron & 6.105 \\
                      positron & 8.574 \\
                               &        \\
       \end{tabular}
       &
       17.5
       \\
   \hline
   \hline
   \end{tabular}
   \label{Tabl_2}
\end{table}

The performed numerical experiments have confirmed the data of the analytical studies, described above, on the feasibility of simultaneous acceleration and focusing of both electron and positron bunches in one and the same acceleration structure without changing its parameters (except the delay time of the accelerated bunch). Also the possibility of generating the acceleration wakefield of the same amplitude by using a regular drive-bunch sequence having the same total charge is shown. In this case, as opposed to the event with a single drive bunch, we have the improvement in the drive bunch transportation, since the transverse force acting on individual bunches becomes either focusing, or essentially smaller in the amplitude.  The results of numerical studies are presented in the figures given below.
\begin{figure}[!th]
  \centering
  \includegraphics[width=0.49\textwidth]{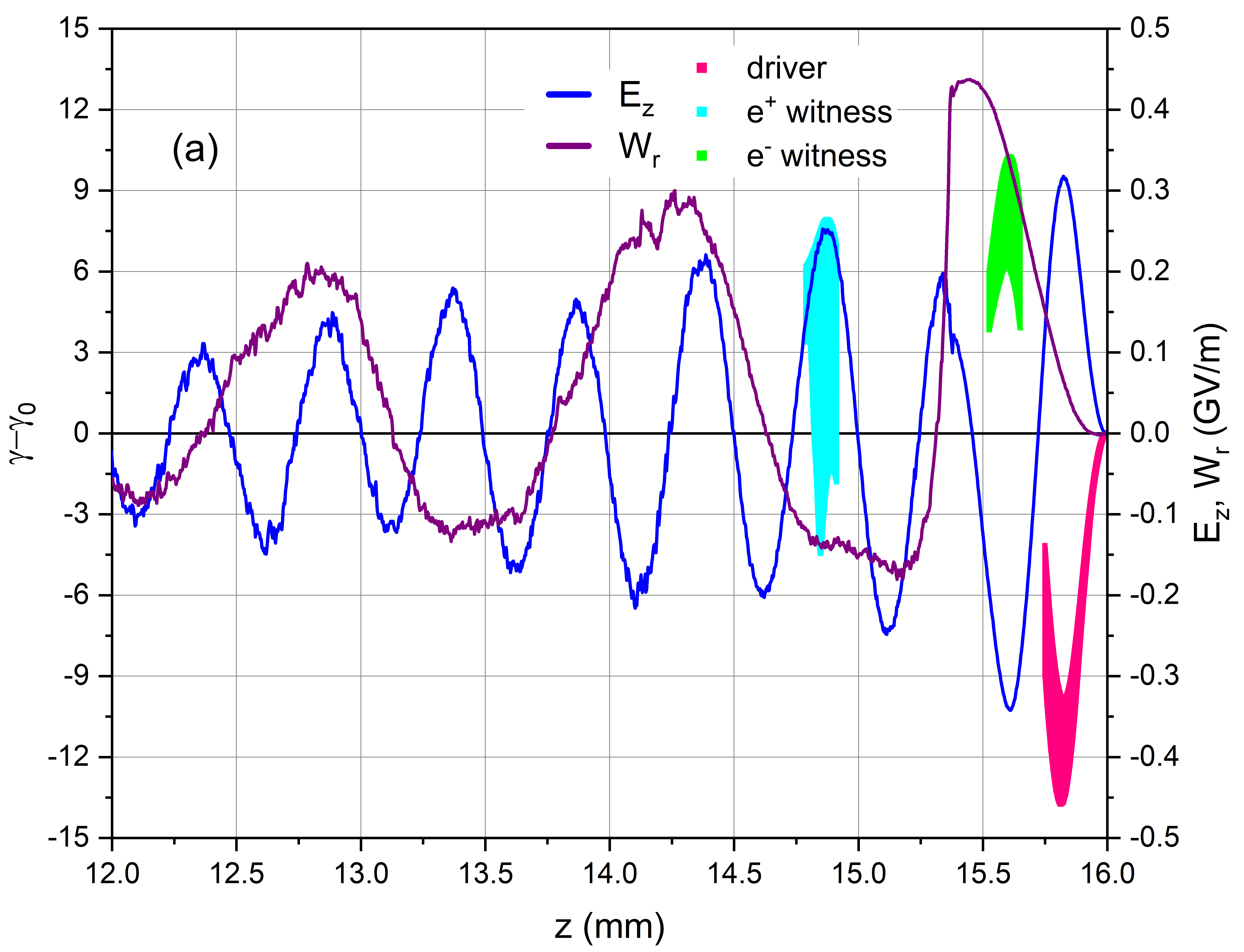}
   \includegraphics[width=0.49\textwidth]{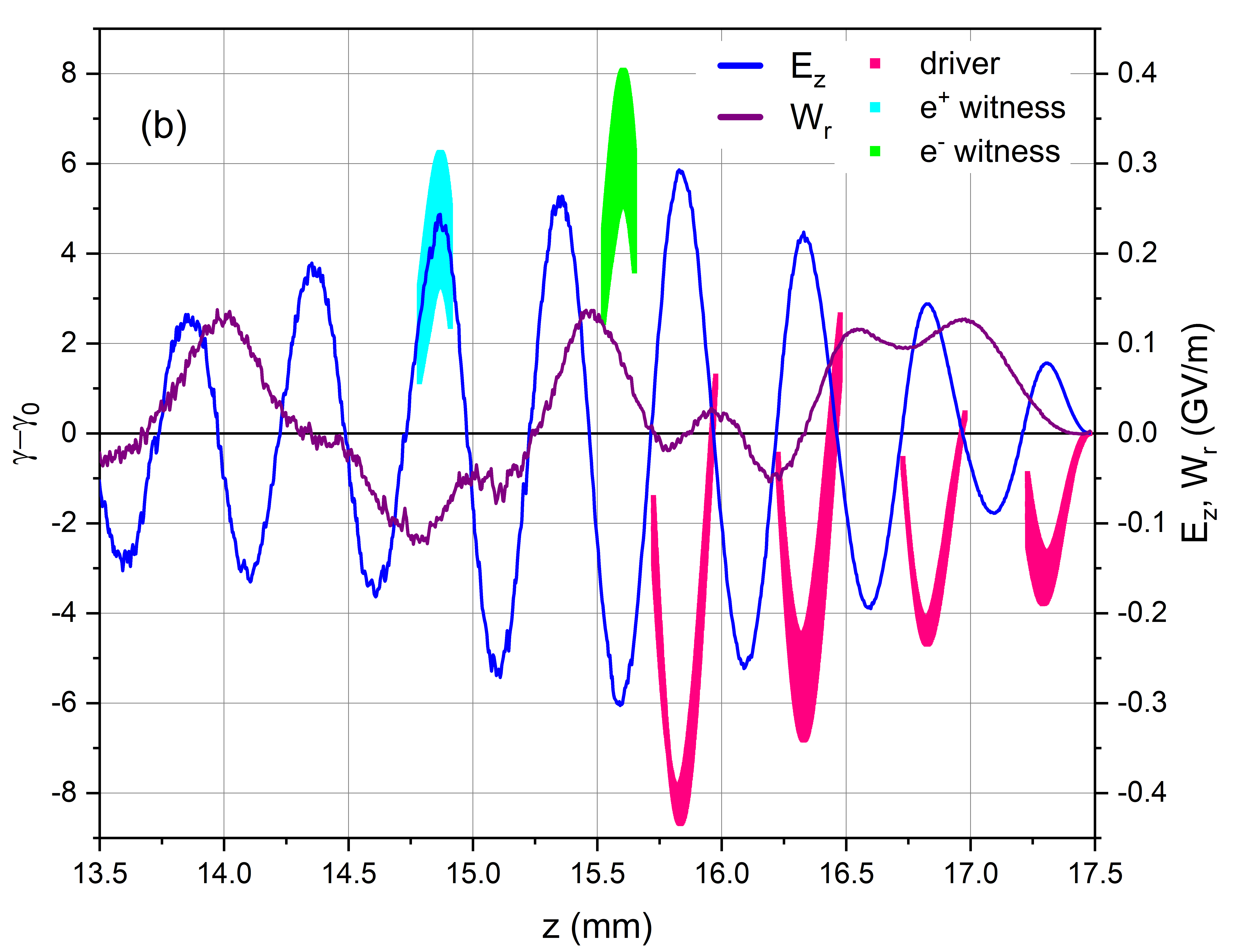}
  \caption{Lorentz-factors of drive-bunch particles (magenta dots) and witness-bunch particles (green dots for electrons and cyan dots for positrons), combined with the longitudinal structure of the amplitudes of axial ($E_z(z)$, blue) and radial  ($W_r(z)$, purple) wakefields at the point of time when the head of the first bunch has reached the end of the  waveguide section. Here $\gamma_0$ is the initial relativistic factor of the injected particles at $z=0$.  The wakefields were computed for the transverse coordinate that corresponded to the initial bunch radius ($r=250\,\rm \mu m$).  The functions in subfig.~(a) were obtained for the case with the single drive bunch, in subfig.~(b) -- for the resonant four-drive bunch sequence. Both the electron and positron bunches are simultaneously in the accelerating and focusing field with the use of both the single electron bunch and their sequence as drivers. In both cases, the rate of witness-bunch acceleration is nearly the same.}\label{Fig:09}
\end{figure}

\begin{figure}[!tbh]
  \centering
  \includegraphics[width=0.49\textwidth]{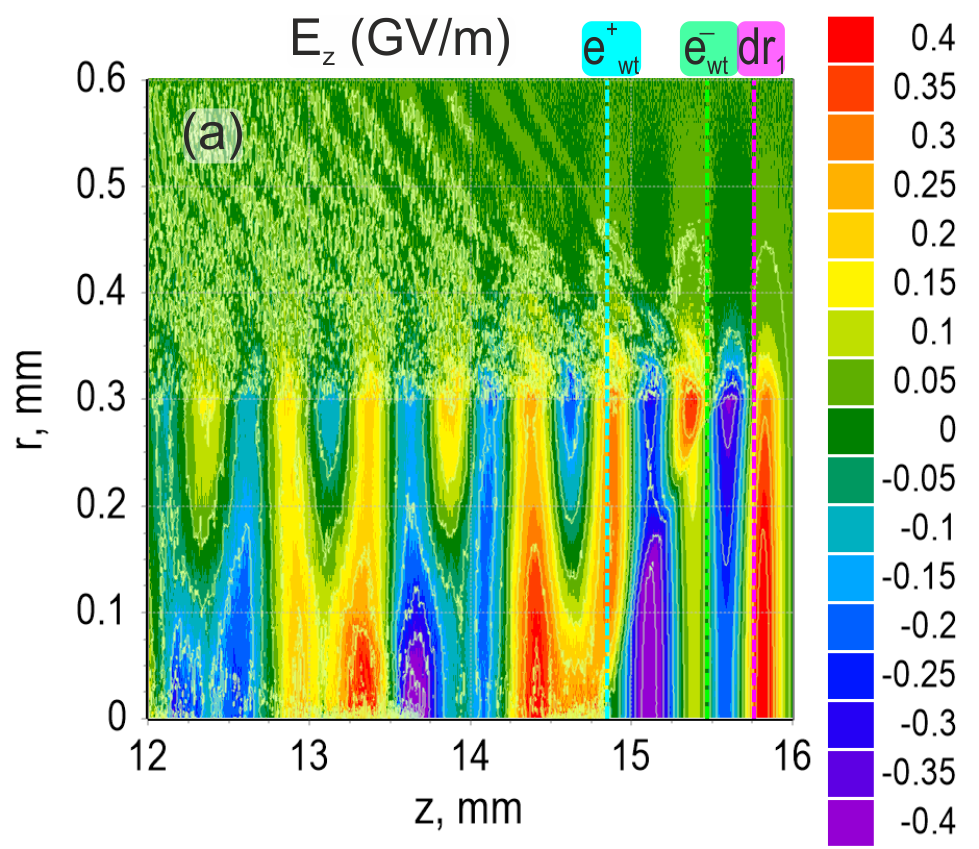}
   \includegraphics[width=0.49\textwidth]{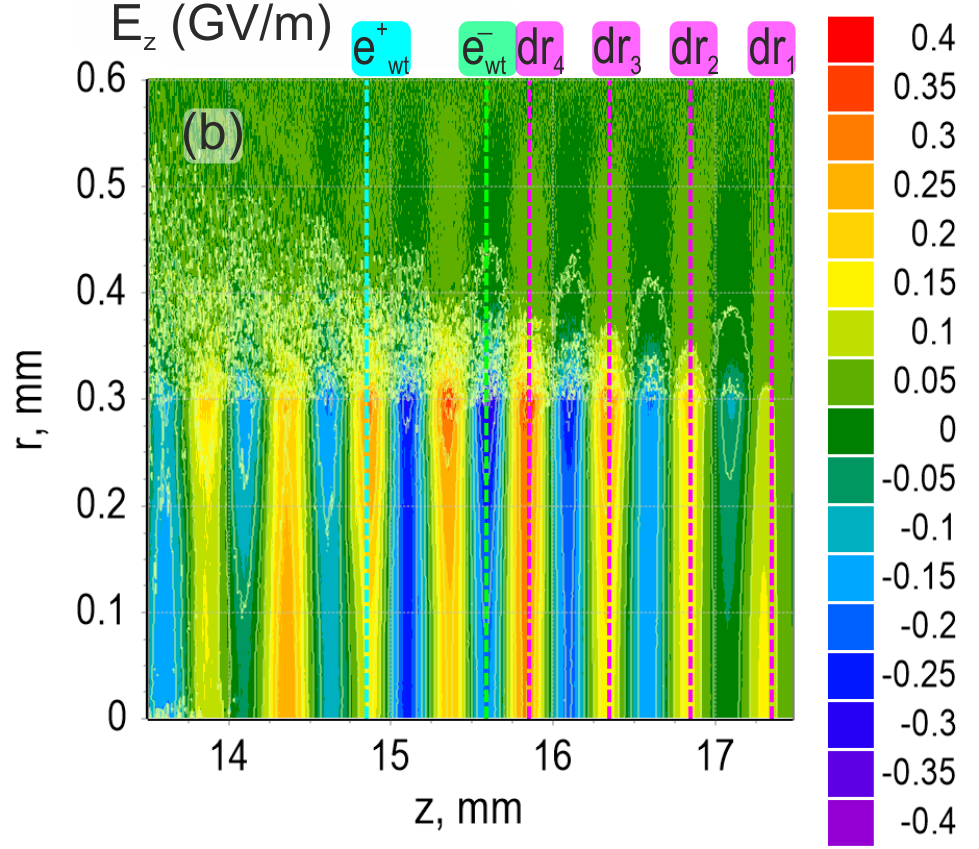}\\
    \includegraphics[width=0.49\textwidth]{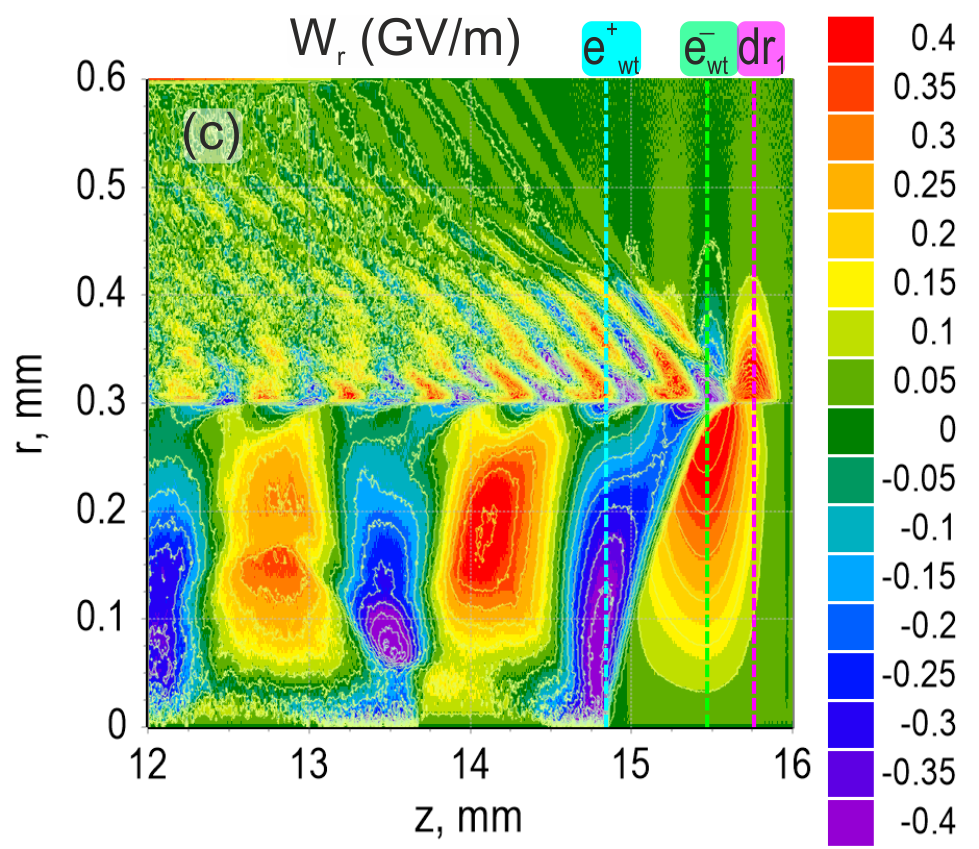}
   \includegraphics[width=0.49\textwidth]{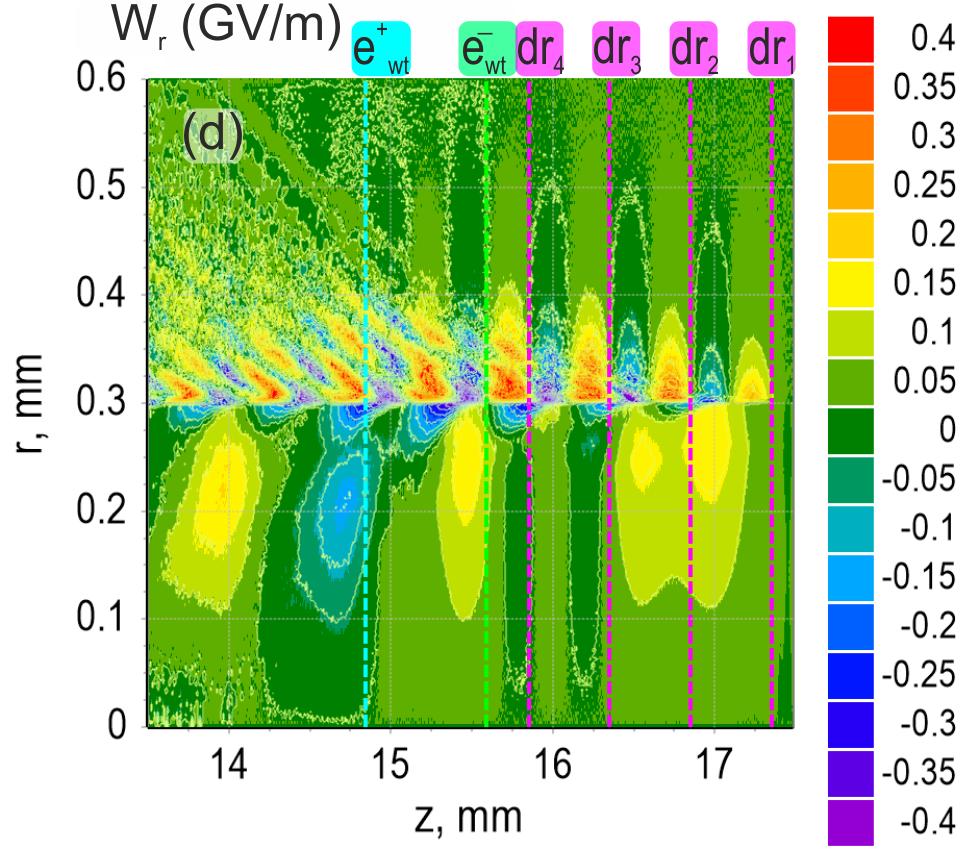}
   \caption{Color map of the axial $E_z(r,\,z)$ and radial $W_r(r,\,z)$ driver bunch-excited wakefields. Vertical dashed lines with their captions mark the middle positions of the witness bunches (cyan color for positrons and green color for electrons) and driver bunches (magenta). Figures (a) and (c) were obtained for a single drive bunch, and figures (b) and (c) -- for the resonant four drive-bunch sequence.}\label{Fig:12}
\end{figure}

Figure ~\ref{Fig:09} shows the longitudinal structure of the axial $E_z(z)$ and radial $W_r(z)$ drive bunch-excited wakefields, calculated at $r=250\, \rm\mu m$, which corresponds to the initial bunch radius. Figure~\ref{Fig:09}a corresponds to the case with the single bunch of charge $-2\, nC$, used as a driver and Fig.~\ref{Fig:09}b illustrates the case, where the driver is represented as a regular sequence of equally charged $-0.5\, nC$ four bunches separated by the injection period  $T_r$. The same plots show the phase planes of the drive and witness bunch particles
(the change in the relativistic factor $\gamma-\gamma_0$ vs. the longitudinal coordinate $z$)
at the time moment when the driver head reaches the end of the acceleration structure. The wakefield distributions presented in Fig.~\ref{Fig:09} and in the following figures were obtained with the use of the positron witness-bunch. However, in view of the smallness of the witness-bunch charge in comparison with the single drive bunch charge and the total charge of the drive bunch sequence, the wakefield distribution remains practically the same in the case of the electron witness-bunch. All particles in a single drive bunch (Fig. ~\ref{Fig:09}a) lose their energy, transferring it to the wakefield and plasma particles, while the electron witness bunch increases its energy. However, the positron witness not only gains, but also looses a small part of its energy, this being due to the distortion of the plasma column density, and the change in the transverse distribution of the wakefield, particularly this being valid for paraxial positrons (see below Fig.~\ref{Fig:12}a). The maximum rate of acceleration, calculated as the energy gain divided by the distance covered, is equal to $335.8\, \rm MeV/m$ for electrons,  and to $272\, \rm MeV/m$ for positrons. These values correlate with the respective peak values of the acceleration fields. In case of the resonant drive sequence, all particles of the first drive bunch give up energy to the system, and the rest bunches of the drive sequence as a whole give up energy, but there is an insignificant number of particles in the bunch heads, which increase their energy by taking away energy from the excited field. The latter, though, has only a slight effect on the wakefield amplitude, which insignificantly decreases in comparison with the field amplitude in the single driver case. All particles of both the electron and positron witness-bunches in the generated wakefield show a rise in their energy,
which is caused by a more homogeneous transverse distribution of the axial electric field (see below Fig.~\ref{Fig:12}b).
And a more homogeneous transverse profile in the case of a regular bunch sequence is related to monochromatization of the axial wakefield by the regular bunch sequence. The rate of witness acceleration in the case of the drive bunch sequence is somewhat lower than in the single bunch case, and equals  $265\, \rm MeV/m$ and $215\, \rm MeV/m$  for electrons and positrons, respectively.

As it follows from Fig.~\ref{Fig:09}a and Fig.~\ref{Fig:09}b, the accelerating field for the witness-bunches can act simultaneously as a focusing field.
The transverse field inside the single drive bunch acts as focusing, but in the sequence case, inside the drive bunches, the transverse field shows a more intricate behavior (Fig.~\ref{Fig:09}b).  The first two bunches are in the focusing field, while the third and the fourth bunches are in the alternating-sign wakefield. Though these transverse fields are low in comparison with the amplitude value, the field inhomogeneity along the drive bunch can cause its filamentation. The field inside all the bunches under discussion can be made focusing through increasing the ratio of the annular plasma density (2) to the plasma column density (1).
\begin{figure}[!th]
  \centering
  \includegraphics[width=0.49\textwidth]{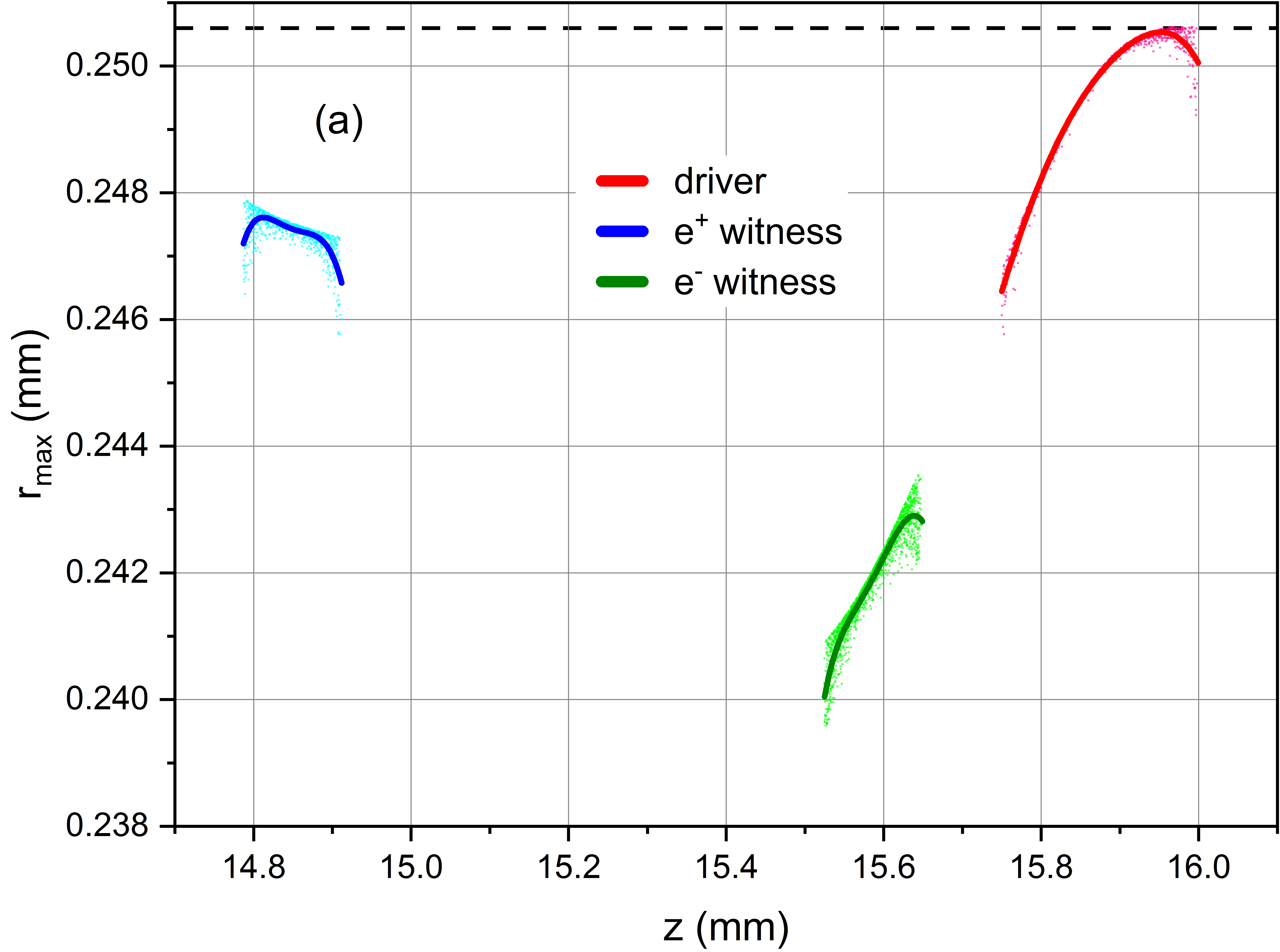}
   \includegraphics[width=0.49\textwidth]{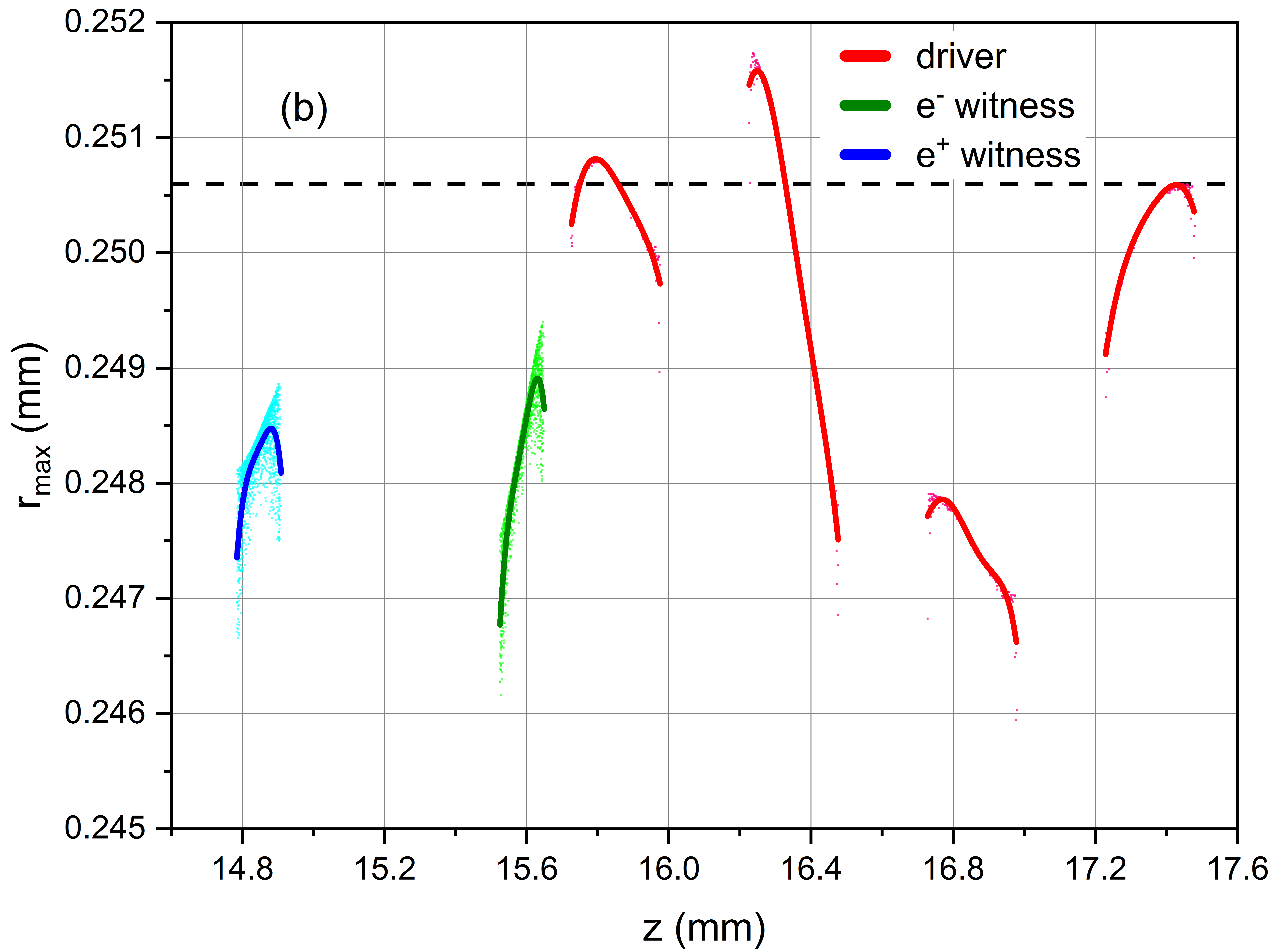}
  \caption{Configurational spaces of bunch particle positions, located at the periphery, and thus forming the bunch radii $r_{max}$ of the drivers (magenta dots) and witnesses (green dots for electron and cyan dots for positron) as functions of the longitudinal coordinate $z$.  The solid lines are the 5th order polynomial fit of the periphery particles radii $r_{max}(z)$. The dashed line denotes the initial radius of the bunches.  The functions shown in fig.~(a) were obtained for a single drive bunch, in fig.~(b) -- for the resonant sequence comprising four drive bunches. The electron and positron witness bunches, and also, the first two bunches of the driver sequence are focused.}\label{Fig:10}
\end{figure}

Figure.~\ref{Fig:10} shows the configurational spaces ($r-z$) of bunch particle positions, located at the periphery, and thus forming  the bunch radii $r_{max}$ as functions of the longitudinal coordinate $z$. From the figure~\ref{Fig:10}a, which displays  the single drive bunch case, it is evident that its tail gets focused through changing the transverse size of the bunch. The electron and positron witnesses are focused all along their whole length, even if inhomogeneously. In the process, the electron witness focusing is essentially greater (maximum compression being $\approx 11\, \rm\mu m$), than that of the positron witness (highest compression is  $\approx 5\, \rm\mu m$). This is due to nearly a two-fold excess of the radial wakefield $W_r(z)$, which acts on the electron witness in comparison with the field focusing the positron witness. As regards the sequence of drive bunches following with the period of the resonant eigenmode $TM_{01}$ of the layered waveguide (Fig.~\ref{Fig:10}b), it is evident that the first drive bunch is focused in the same way as in the case of a single driver, but with a lower focusing efficiency because of  a four times lower charge of each bunch of the sequence. The particles of the 2-nd drive bunch of the sequence are focused along their whole length, even if inhomogeneously, since they are in the region of positive value of the radial wakefield  of different value. The head of the 3-rd drive bunch is focused, but its tail gets defocused.  The behavior of the last 4-th bunch of the sequence is similar, though to a lesser degree. The mentioned transverse dynamics of the 3-rd and 4-th bunches is related to the change of $W_r$ sign along the bunch (see Fig.~\ref{Fig:09}a). Thus, the transverse dimension of the 3-rd and 4-th bunches increases, although the defocusing at the given acceleration distances is rather moderate. The electron and positron witness bunches get fully focused, but this focusing varies along their lengths, because they are found in length-nonuniform focusing radial wakefields (positive and negative for the electron and positron bunch, respectively). Owing to the close values of the transverse force $e^\pm W_r$ acting on the electron and positron witness bunches, the last ones are focused nearly to the same extent. Note that for the bunch drive sequence, the focusing of witness bunches appears less stronger as compared to the single drive bunch case (maximum compression being $4.6\, \rm\mu m$ for electrons and  $4.0\, \rm\mu m$ for positrons). This is caused by different periods of the focusing and accelerating fields. The bunch sequence is in resonance with the accelerating mode $TM_{01}$ of the wakefield,  but is nonresonant with the focusing Langmuir wave, which has a threefold  period than the mode $TM_{01}$ .

\begin{figure}[!th]
  \centering
  \includegraphics[width=0.49\textwidth]{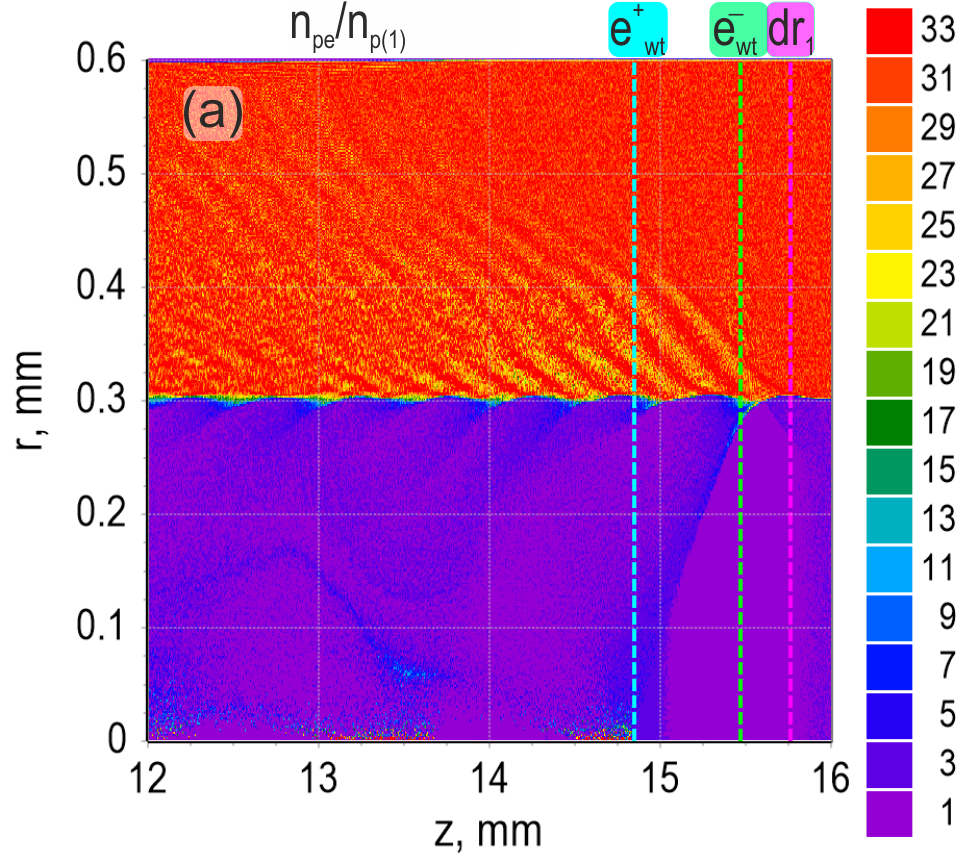}
   \includegraphics[width=0.49\textwidth]{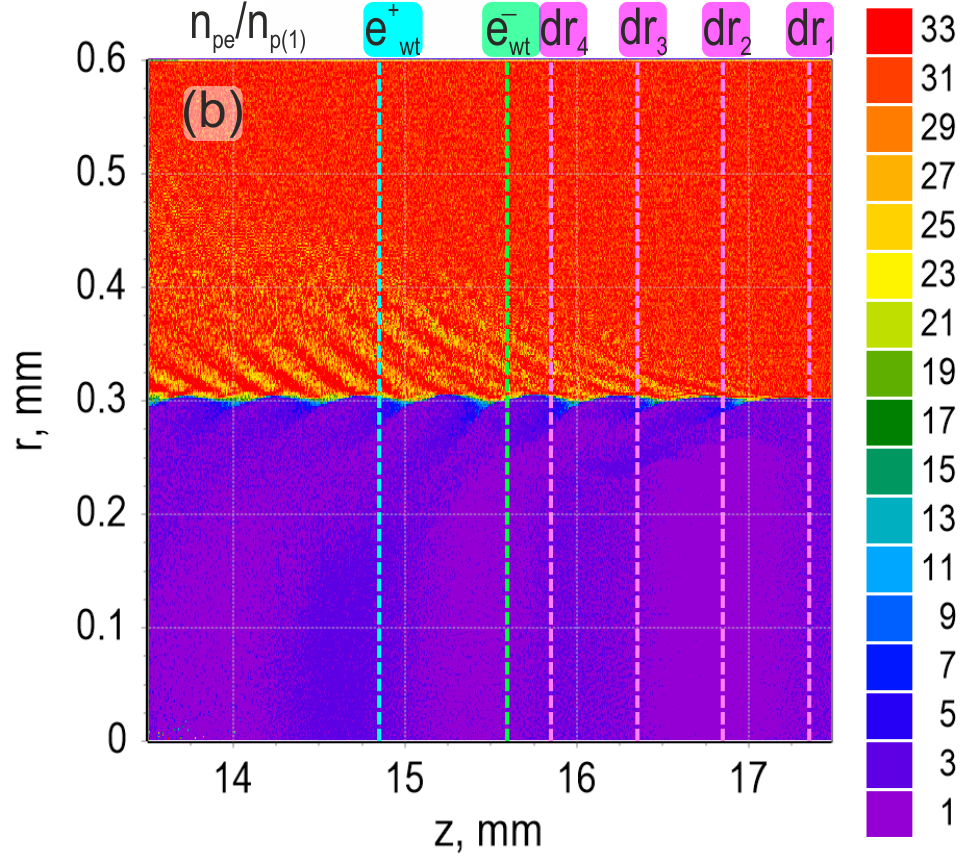}
  \caption{Color map for electron density of the plasma filling the layered waveguide $n_{pe}$, normalized to the initial density of the   plasma column $n_{p(1)}$. Vertical dashed lines with their captions mark the middle positions of the witness bunches (cyan color for positrons and green color for electrons) and drive bunches (magenta). Fig.~(a ) shows the single drive bunch case, fig.~(b) the resonant sequence case involving four drive bunches. The plasma density distribution exhibits two space periods (along z axis), the boundary between two plasma layers distinctly shows the period $0.5\, \rm mm$, corresponding to the period of the  $TM_{01}$ mode.}\label{Fig:11}
\end{figure}

Figure.~\ref{Fig:11} demonstrates the density distribution of plasma electrons filling the layered structure $n_{pe}$. The vertical dashed lines show the  positions of witness/drive bunch centers. At single drive bunch injection into the structure (Fig.~\ref{Fig:11}a), at the boundary between the plasma column (lower density) and the tubular plasma (higher density), there arises the plasma electron density modulation with period $\approx 0.5\, \rm mm$, corresponding to the resonance period of the surface wave  $TM_{01}$ of the layered waveguide. Behind the drive bunch, the electrons from the tubular plasma region bear down into the plasma column (region (1)). For the time delay of the electron witness $t_{del}^{e^-}$ the electrons from this region don't have time to reach the structure axis and the electron density distortion in region (1) inside the witness bunch appears to be small, therefore the wakefield within it is not much different from linear values. However, in the case of the positron witness bunch, the delay time $t_{del}^{e^+}$ of which is  more than  $t_{del}^{e^-}$, the electrons from region (2) do reach the structure axis this time with the result that  the electron density in region (1) gets appreciably distorted, and this, in turn, strongly affects the regularity of the wakefield acting on the accelerated positron bunch. A similar pattern of plasma density modulation is also takes place when injecting the resonant sequence of drivers (Fig.~\ref{Fig:11}b) into the structure. However, for the delay time of the positron witness $t_{del}^{e^+}$, the electrons from region (2) have no time to reach the structure axis, and the electron density distortion of the plasma column appears somewhat lower than in the single bunch case. This, in turn, leads to the fact that the wakefield excited by the resonant driver sequence has a more regular character.

Figure~\ref{Fig:12} shows the color map of the 2D-distribution  of the axial $E_z(r,\,z)$ and radial $W_r(r,\,z)$ drive bunch-excited wakefields in the LPW. Figures~\ref{Fig:12}a and~\ref{Fig:12}c correspond to the single drive bunch injection. From the comparison of these distributions with similar 2D-distributions of the wakefields defined in the analytical calculations, it is evident that the PIC simulation data display a more inhomogeneous field distribution in the both plasma layers. Besides this, at long distances from the drive bunches, the growth of the wakefield period occurs. The dissimilarity of the numerical simulation data from the linear theory results is related to a self-consistent consideration of the nonlinear plasma response and the group velocity dispersion of the excited wave propagation~\cite{Bal2001JETP}. The change in the axial field structure causes the slope of level lines $W_r$ inside the positron bunch, and this, in turn, will lead to the change in the bunch particle distribution in the radial direction. This effect may turn out to be essential for long acceleration lengths. In the case of resonant driver sequence injection into the LPW, the axial wakefield  $E_z(r,\,z)$ (Fig.~\ref{Fig:12}b) exhibits the coordinate dependence close to that derived in the linear theory (compare with Fig.~\ref{Fig:04}a). As regards the radial wakefield $W_r(r,\,z)$, here we have, just as in the case of a single drive bunch, an appreciable distortion of the field as compared to the linear theory. A more significant change in the transverse wakefield is due to the fact that it is determined by the plasma of essentially lower density, therefore, the nonlinear effects of plasma response are of  great importance.

\begin{figure}[!th]
  \centering
  \includegraphics[width=0.49\textwidth]{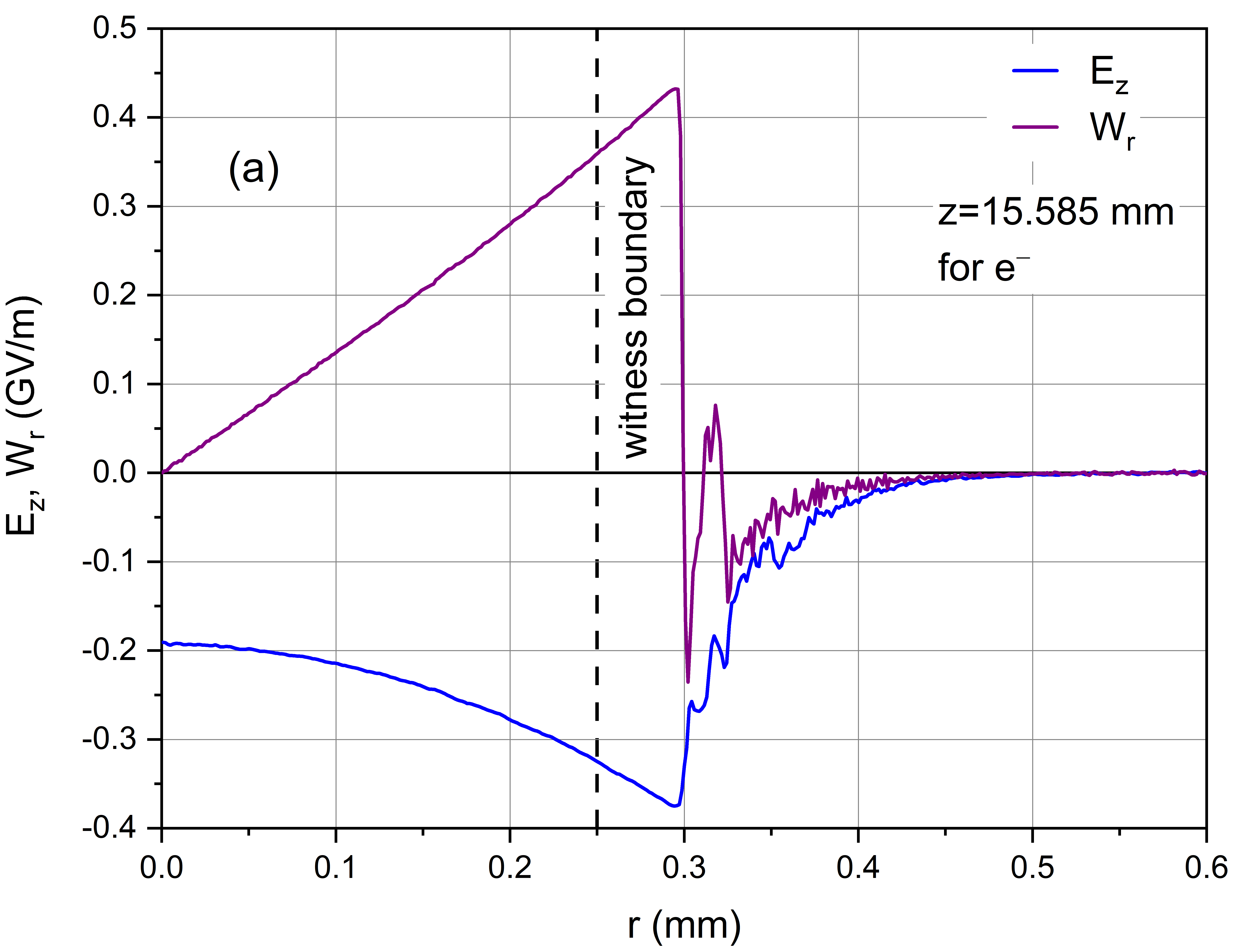}
   \includegraphics[width=0.49\textwidth]{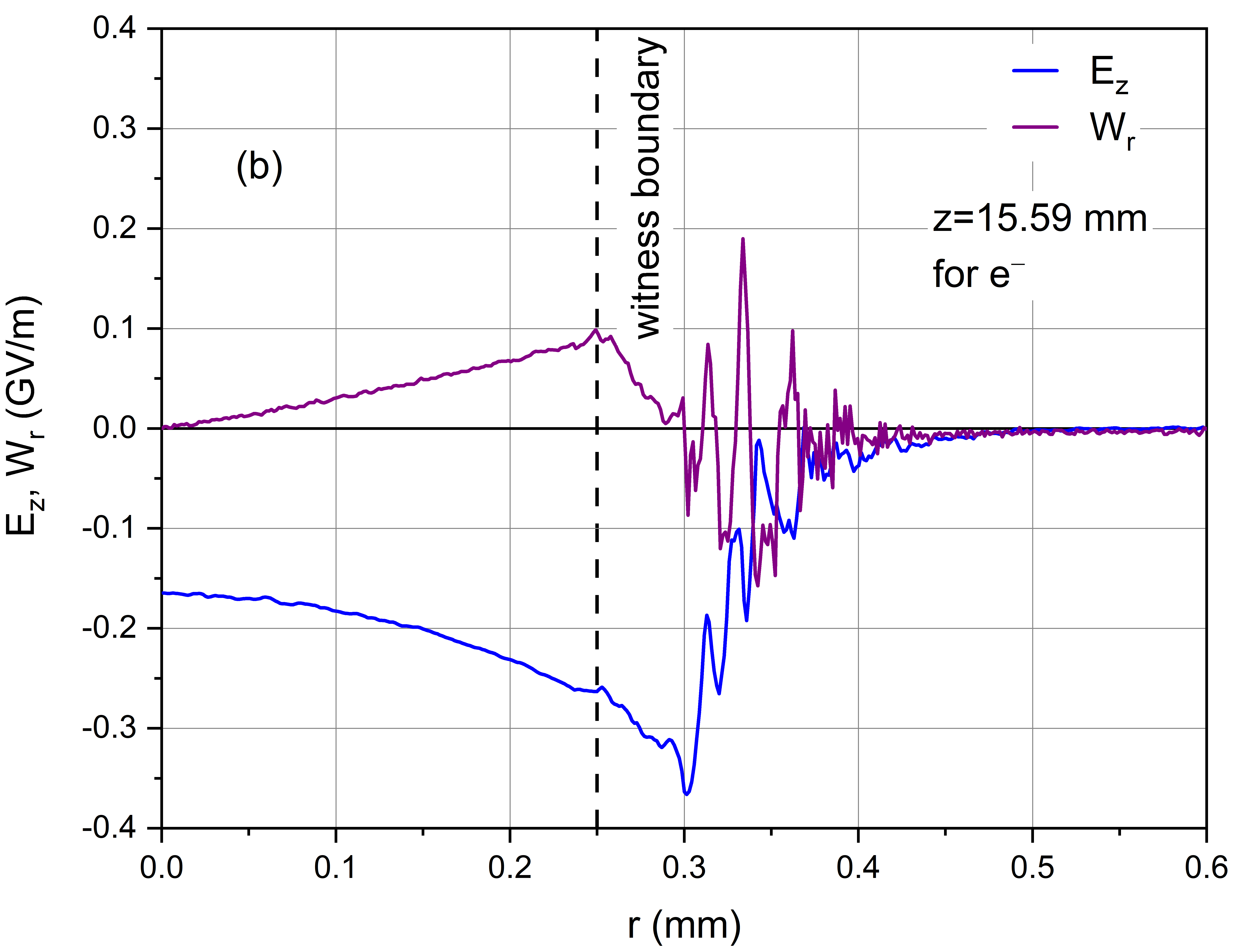}
  \caption{Transverse profile of the axial $E_z(r)$ and radial $W_r(r)$ wakefields acting on the particles of the electron witness bunch along its middle part (i.e., along green dotted line in  Fig.~\ref{Fig:12}). Fig.~(a)  shows the single drive bunch case, fig.~(b) the resonant 4-drive bunch sequence case. The transverse wakefield has a linear dependence on the radial coordinate.}\label{Fig:13}
\end{figure}

Figure~\ref{Fig:13}  shows the radial profile of the axial $E_z(r)$ and transverse $W_r(r)$ wakefields, calculated in the middle of the electron witness bunch. As it follows from the given dependencies, in the cases of both the single drive bunch and the sequence of drive bunches, the profile of transverse fields acting on the electron witness is qualitatively the same. The axial wakefield $E_z$ shows the surface character as it increases quadratically by a factor of $1.5-2$ from the axis to the lateral surface of the electron bunch ($r=250.6\,\rm \mu m$). Owing to this inhomogeneity of the longitudinal electric field, the paraxial electrons get a higher acceleration than at the bunch periphery, resulting in a particle energy spread of the bunch. The radial wakefield $W_r$ exhibits practically a linear dependence, increasing up to $0.43\, \rm GV/m$ at single drive injection, and up to  $0.1\,\rm GV/m$ at injection of the resonant 4-drive bunch sequence. The linear dependence of the radial wakefield $W_r$ precludes achromatic aberration in the process of witness bunch focusing. The wakefield profile, obtained from the PIC simulation, is in qualitative agreement with the analytical calculations, and is in quantitative  agreement with the results presented in section~\ref{sec:3}.

\section{Summary}~\label{sec:5}
Analytical studies and preliminary PIC simulation confirm that layered  plasma waveguide can be used as candidate for promising wakefield accelerators, both electon and positron accleration,  based on   excitation  of high-gradient fields by charged bunches. In the linear approximation of plasma dynamics, analytical expressions for the excited longitudinal and transverse wakefields are obtained. The dispersion of the TM-modes of the LPW was obtained and analyzed, and it was shown that there is a single TM wave resonant with the electron bunch.  Based on the obtained analytical expressions, the structures of the amplitudes of the axial and radial wakefields are numerically investigated. The electromagnetic TM wave is a surface wave whose amplitude decreases from the interface between two layers.  It is found that for certain ratios of the densities of the outer and inner plasmas, it is possible both to simultaneously accelerate and focus the drive and witness bunches. The 2.5D particle-in-cell code simulations  show the LPWA  can provide the stable transport both accelerated  electron and positron bunches. We carried our the studies for short plasma layered wakefiled  accelerator.  Additional simulation is required at a longer acceleration distance to obtain more information on the quality of acceleration using LPW (bunch emittance, energy distribution, efficiency) and to compare the characteristics of accelerated bunches with those already obtained in other new acceleration methods.

\begin{acknowledgments}
The study is supported by the National Research Foundation of Ukraine under the program “Excellent Science in Ukraine” (project No. 2023.03/0182).
\end{acknowledgments}


\bibliographystyle{elsarticle-num}
\bibliography{Bibliography_LPWA}

\end{document}